\documentclass[useAMS,usenatbib]{mn2e}

\usepackage{natbib}
\usepackage{gensymb}

\usepackage{latexsym,graphicx}
\usepackage{color}
\usepackage{fixltx2e}
\usepackage{verbatim} \usepackage{float}
\usepackage{amsmath,amssymb}
\usepackage{times}
\usepackage{siunitx}
\newcommand{\mpo}{\textcolor{black}}
\newcommand{\mpon}{\textcolor{black}}
\newcommand{\dmam}{\textcolor{black}}

\usepackage{setspace}
\usepackage{rotating}
\usepackage{pdflscape}
\usepackage{subfig}

\newcommand\hii{H\,{\sc ii} \,}

\def\apgt{\ {\raise-.5ex\hbox{$\buildrel>\over\sim$}}\ }
\def\aplt{\ {\raise-.5ex\hbox{$\buildrel<\over\sim$}}\ }

\let\oldhat\hat
\renewcommand{\hat}[1]{\oldhat{\mathbf{#1}}}

\title[Nebulae and remnants of very massive stars]{Wind nebulae and supernova remnants of very massive stars}

\author[D. M.-A.~Meyer et al.]
       {D. M.-A.~Meyer\thanks{E-mail: dmameyer.astro@gmail.com}$^{1}$, M.~Petrov$^{2}$  and M.~Pohl$^{1,3}$ \\
       $^{1}$ Institut f\" ur Physik und Astronomie, Universit\" at Potsdam, Karl-Liebknecht-Strasse 24/25, 14476 Potsdam, Germany\\
       $^{2}$ Max Planck Computing and Data Facility (MPCDF), Gießenbachstrasse 2, D-85748 Garching, Germany\\ 
       $^{3}$ DESY Platanenallee 6, D-15738 Zeuthen, Germany \\        
       }

\begin{document}

\date{Received; accepted}

\maketitle

\label{firstpage}

\begin{abstract} 
\textcolor{black}{
A very small fraction of (runaway) massive stars have masses exceeding $60$-$70\, \rm M_{\odot}$ and are predicted to evolve as Luminous-Blue-Variable and Wolf-Rayet stars before ending their lives as core-collapse supernovae. 
Our 2D axisymmetric hydrodynamical simulations explore how a fast wind ($2000\, \rm km\, \rm s^{-1}$) and high mass-loss rate ($10^{-5}\, \rm M_{\odot}\, \rm yr^{-1}$) can impact the morphology of the circumstellar medium. It is shaped as 100 pc-scale wind nebula which can be pierced by the driving star when it supersonically moves with velocity $20$-$40\, \rm km\, \rm s^{-1}$ through the interstellar medium (ISM) in the Galactic plane.
%
%
The motion of such runaway stars displaces the position of the supernova explosion out of their bow shock nebula, imposing asymmetries to the eventual shock wave expansion and engendering Cygnus-loop-like supernova remnants. 
We conclude that the size (up to more than $200\, \rm pc$) of the filamentary wind cavity in 
which the chemically enriched supernova ejecta expand, mixing efficiently the wind and ISM materials by at least $10\%$ in number density, can be used as a tracer of the runaway nature of the very massive progenitors of such $0.1\, \rm Myr$ old remnants. 
Our results motivate further observational campaigns devoted to the bow shock of the very massive stars BD+43 3654 and to the close surroundings of the synchrotron-emitting Wolf-Rayet shell G2.4+1.4. 
}
\end{abstract}



\section{Introduction}
\label{sect:introduction}

Massive stars ($M_{\star}\ge8\, \rm M_{\odot}$) are objects whose formation is an uncommon 
but crucial event in the interstellar medium (ISM) of our Galaxy~\citep{langer_araa_50_2012}. 
Their strong winds release a large amount of momentum and energy into their surrounding ISM and form bubbly circumstellar 
structures which chemically enrich their local environment~\citep{weaver_apj_218_1977}. 
Wind bubbles of main-sequence, OB-type massive stars are predicted to expand up to $\simeq 100\, \rm pc$ 
away from the star. The shocked photoionised stellar wind material and the ISM gas are separated by an 
unstable contact discontinuity segregating the inner hot, diluted shocked wind gas from 
the outer layer of cold, dense shocked ISM gas. 
These nebulae grow inside the huge \hii regions generated by the ionising radiation field of early-type hot massive 
stars~\citep{dyson_ass_35_1975,dyson_apss_37_1975}, however, sufficiently massive stellar wind bubbles can trap the ionisation 
front of their own central star~\citep{Dwarkadas_HEDP_2013}. 
Wind bubbles mainly radiate by means of optical forbidden-line emission~\citep{schneps_apj_246_1981,smith_mnras_211_1984}, 
in the X-ray energy band via thermal photons emitted by the inner hot region~\citep{zhekov_mnras_443_2014}, and in the infrared 
waveband by reprocessing of starlight by ISM dust that is trapped into the outer layer of the bubble~\citep{vanburen_apj_329_1988}. 
Stellar wind bubbles of static massive stars conserve their spherical shape regardless of the central star's evolution, 
see~\citet{garciasegura_1996_aa_305,freyer_apj_594_2003,freyer_apj_638_2006,dwarkadas_apj_667_2007}.

\begin{figure}
        \centering
                \includegraphics[width=1.0\columnwidth]{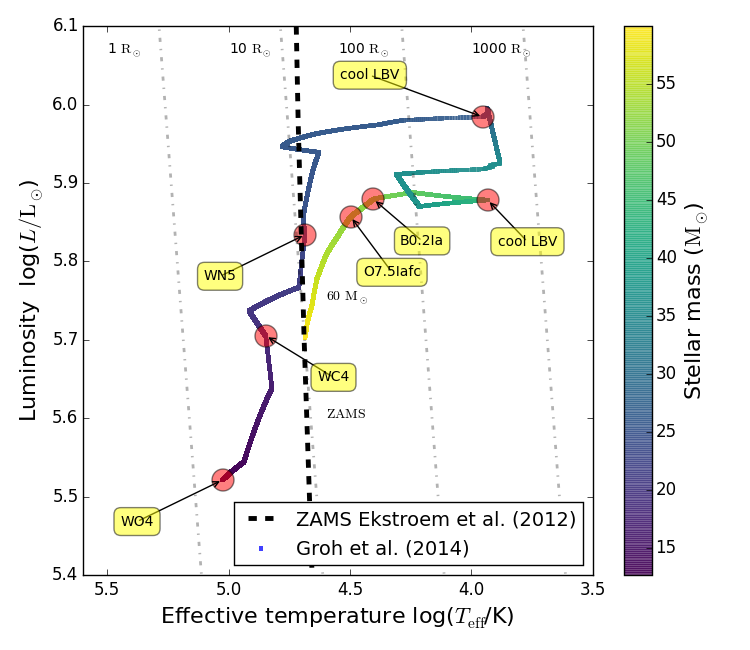}
        \caption{ 
        	 Herzsprung-Russel diagram of the non-rotating $60\, \rm M_{\odot}$ zero-age-main-sequence 
        	 (ZAMS) star \dmam{considered in this study}, based on~\citet{groh_aa564_2014}. 
               }      
        \label{fig:hdr}  
\end{figure}

The spherical symmetry of stellar wind nebulae is 
broken if their central star moves with respect to 
their local ambient medium, and they eventually turn into a similarly organised but arc-like shape~\citep{baranov_sphd_15_1971,wilkin_459_apj_1996}. 
The resulting so-called stellar wind bow shocks form around the 4 to 10 per cent of all main-sequence stars that move 
supersonically through the ISM~\citep[see][]{gies_apjs_64_1987,blau1993ASPC...35..207B, huthoff_aa_383_2002}. 
The physics of bow shocks around OB stars partly relies of the (in)efficiency of heat conduction~\citep{meyer_mnras_464_2017,2019A&A...625A...4G} while their 
detailed internal structure depends on the coupling between dust and gas~\citep{henney_mnras_486_2019,henney_486_mnras_2019,henney_2019_arXiv190400343H}. 
Their morphology is a function of the stellar mass-loss rate~\citep{gull_apj_230_1979,Gvaramadze_2013,meyer_2014a}, 
ambient medium density~\citep{kaper_apj_475_1997}, and stellar bulk motion~\citep{meyer_2014a}, respectively. 
%
%
The evolution of massive stars induces brutal changes in the stellar wind properties~\citep{langer_araa_50_2012}. The dense but slow wind of 
evolved red supergiant stars makes bow shocks 
prone to develop thin-shell instabilities~\citep{dgani_apr_461_1996,blondin_na_57_1998,mackey_apjlett_751_2012,meyer_2014bb}, 
which can be inhibited either by an external ionising radiation field~\citep{meyer_2014a} or by the ambient magnetisation of the ISM~\citep{vanmarle_aa_561_2014}. 
These dusty nebulae are mainly observable in the infrared waveband~\citep{vanburen_aspc_35_1993,vanburen_aj_110_1995,vanmarle_apj_734_2011,vanmarle_aa_537_2012,peri_aa_538_2012,2014Natur.512..282M,peri_aa_578_2015,kobulnicky_apjs_227_2016,kobulnicky_aj_154_2017,kobulnicky_apj_856_2018} although they thermally emit in the other wavebands, see~\citet{kaper_apj_475_1997,jorissen_532_aa_2011}.  
Interestingly, such bow shocks are suspected to accelerate protons and electrons in the stellar wind to high energies and to act as cosmic ray injectors \citep{delvalle_aa_550_2013,delvalle_mnras_448_2015,delvalle_apj_864_2018}, producing variable high-energy and 
$\gamma$-ray emission~\citep{delvalle_aa_563_2014}, however at much lower luminosity than, e.g. pulsars or supernova remnants~\citep{DeBecker_mnras_471_2017,tola_apj_838_2017,binder_aj_157_2019}. 
Recent measures have reported synchrotron emission from the Wolf-Rayet ring bubble G2.4+1.4~\citep{2019arXiv190912332P}.

When bow-shock-driving massive stars cease to evolve, the distorted bubble nebula shaped throughout the  
post-ZAMS (zero-age-main-sequence) star life constitutes the environment in which the moving high-mass star 
dies, e.g. as a  core-collapse supernova~\citep{franco_pasp_103_1991,rozyczka_mnras_261_1993,ekstroem_aa_537_2012}. 
The deviations from the spherically-symmetric expansion of the shock wave \mpo{depend on} the density and mass 
accumulated in the circumstellar medium of the progenitor. 
The blastwave first expands in a cavity of unshocked stellar wind, before interacting with the reverse shock 
of the bow shock, whose presence slows down the progression of the shock wave in the direction of motion of the 
progenitor, while it faciliates its rapid expansion in the opposite direction, inducing characteristic asymmetries in the 
supernova remnant.  
%
%
Typical examples \mpo{among 
young supernova remnants} are Kepler's supernova remnant~\citep{borkowski_apj_400_1992,velazquez_apj_649_2006,chiotellis_aa_537_2012}, 
Tycho~\citep{vigh_apj_727_2011,williams_apj_770_2013}, or the Cygnus Loop~\citep{meyer_mnras_450_2015,fang_mnras_464_2017}. Note that asymmetries can be caused by the geometry of the explosion itself~\citep{toledo_mnras_442_2014}. 
\mpo{Having established that the morphology of supernova remnants from massive progenitors is a function of their} past stellar evolution,  
the question \mpo{arises, which massive stars are }
most prone to generate a dense asymmetric circumstellar medium and hence 
the most aspherical remnants?

\begin{figure}
        \centering
                \includegraphics[width=1.0\columnwidth]{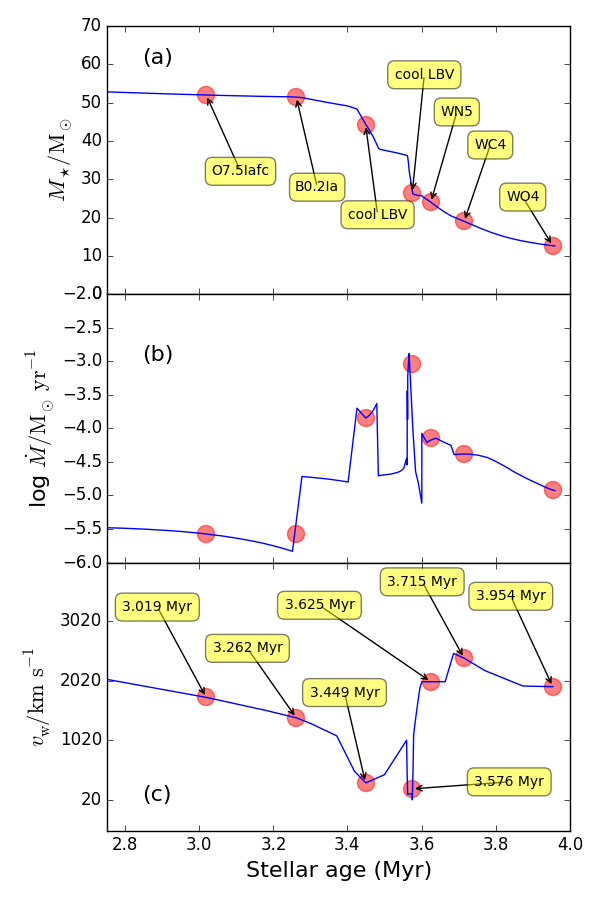}
        \caption{ 
        	 Post-main-sequence evolution of the stellar mass (top, panel a), mass-loss rate (middle, panel b), and 
        	 wind velocity (bottom, panel c) as a function of time (in Myr), that we used as \mpo{initial} 
        	 condition in our hydrodynamical simulations. 
                 }      
        \label{fig:star}  
\end{figure}

\begin{table*}
	\centering
	\caption{
	Simulation models in this study. The table gives the space velocity of the moving star, $v_{\star}$. 
	The runs are labelled as "CSM" for the pre-supernova modelling of the circumstellar medium and as "SNR" for 
	the simulations of their subsequent remnants, respectively.    
	}
	\begin{tabular}{lcccr}
	\hline
	${\rm {Model}}$  &  $v_{\star}$ ($\rm km\, \rm s^{-1}$)  & Grid size    & Grid mesh  &      description  \\ 
	\hline    
	\textcolor{black}{Run-0-CSM}        &  $0$                                  & $[0;150]\times[-150;150]$           &    $3000 \times 6000\, \mathrm{ cells}$         &      static wind bubble   \\  
	\textcolor{black}{Run-10-CSM}       &  $10$                                 & $[0;150]\times[-200;100]$           &    $2000 \times 4000\, \mathrm{ cells}$         &      wind bubble with off-centered slowly-moving star    \\  
	\textcolor{black}{Run-20-CSM}       &  $20$                                 & $[0;175]\times[-250;100]$           &    $2000 \times 4000\, \mathrm{ cells}$         &      bow shock of runaway star moving with Mach number ${M}=1^{a}$    \\  	
	\textcolor{black}{Run-40-CSM}       &  $40$                                 & $[0;150]\times[-300;100]$           &    $1500 \times 4000\, \mathrm{ cells}$         &      bow shock of runaway star moving with Mach number ${M}=2^{a}$    \\
	\textcolor{black}{Run-0-SNR}        &  $0$                                  & $[0;200]\times[-200;200]$           &    $3500 \times 7000\, \mathrm{ cells}$         &      supernova remnant of static progenitor       \\  	
	\textcolor{black}{Run-10-SNR}       &  $10$                                 & $[0;200]\times[-230;170]$           &    $4000 \times 8000\, \mathrm{ cells}$         &      supernova remnant of slowly-moving progenitor             \\	
	\textcolor{black}{Run-20-SNR}       &  $20$                                 & $[0;200]\times[-275;175]$           &    $3200 \times 7000\, \mathrm{ cells}$         &      supernova remnant of progenitor moving with Mach number ${M}=1^{a}$       \\  	
	\textcolor{black}{Run-40-SNR}       &  $40$                                 & $[0;200]\times[-330;170]$           &    $3200 \times 8000\, \mathrm{ cells}$         &      supernova remnant of progenitor moving with Mach number ${M}=2^{a}$             \\	
	\hline    
	\end{tabular}
\label{tab:models}\\
\footnotesize{(a) The stellar motion is supersonic with respect to the unperturbed ISM sound speed.}
\end{table*}

The accumulation of mass in the surroundings of an evolving star
can arise from the brutal release of dense shells of material \mpo{during the post-}main-sequence phase.  
As additional effect to that of the progenitor's stellar motion, this naturally leads to asymmetric \dmam{and inhomogeneous} 
cavities in which the supernova blastwave will expand.  
This particularly affects high-mass stars evolving through eruptive stellar evolutionary stages such as so-called luminous blue variable or Wolf-Rayet phases~\citep{gonzales_aa_561_2014,humphreys_apj_844_2017}, characterised by the sudden inflation of the \mpo{star} together with an increase of the star's mass-loss rate and wind velocity. This results in the ejection of dense shells of envelope material into the stellar surroundings~\citep{sanyal_aa_597_2017,graefener_aa_608_2017,Grassitelli_aa_614_2018}. 
\textcolor{black}{
Such evolving high-mass stars, that are rare and predicted to end their life as a core-collapse 
supernova~\citep{2005A&A...429..581M,ekstroem_aa_537_2012}, have been observed both in the Milky 
Way~\citep{2001NewAR..45..135V} and in the Large Magellanic Cloud~\citep{2014A&A...565A..27H}. 
Moreover, a significant fraction of them have fast proper motion, which can be explained by many-body gravitational interaction leading to their escape from their parent cluster~\citep{2011MNRAS.410..304G,2013MNRAS.430L..20G}.  
\textcolor{black}{Particularly, this mechanism is able to produce very massive ($\simeq 55$-$85\, \rm M_{\odot}$) 
fast-moving stars such as the main-sequence star BD+43$\degree$3654 which runs away from the Cygnus OB2 region~\citep{comeron_aa_467_2007}.}
Hence, these moving stellar objects that are inclined to provoke morphological distortions of their wind 
nebulae~\citet{ohara_apj_598_2003} are also good candidates for the production of strongly asymmetric supernova remnants. 
}
In this study, we explore the effects of consecutive luminous blue variable and Wolf-Rayet winds onto the shaping of 
nebulae around a $60\, \rm M_{\odot}$ star~\citep{freyer_apj_594_2003,vanmarle_aa_469_2007,2011ApJ...737..100T,2017MNRAS.470.2283W} 
(see Fig.~\ref{fig:hdr}), and we investigate how \dmam{its bulk motion} can affect the development of asymmetries 
in their subsequent supernova remnant in the spirit of the first paper of this series devoted to runaway red 
supergiant progenitors~\citep{meyer_mnras_450_2015}.

Our study is divided as follows. First, we present in Section~\ref{sect:method} the methods for the numerical modelling 
of the circumstellar medium of a moving, massive Wolf-Rayet-evolving star. 
We present our results for the dynamics of the stellar surroundings from the ZAMS to the supernova remnant phase 
in Section~\ref{sect:snr}. Particularly, we concentrate on the mixing of material induced in the remnants and on their 
associated thermal X-ray emission properties. 
Our outcomes are further discussed in Section~\ref{sect:discussion}. Finally, we conclude in Section~\ref{section:cc}.

 \begin{figure*}
        \centering
                \includegraphics[width=1.0\textwidth]{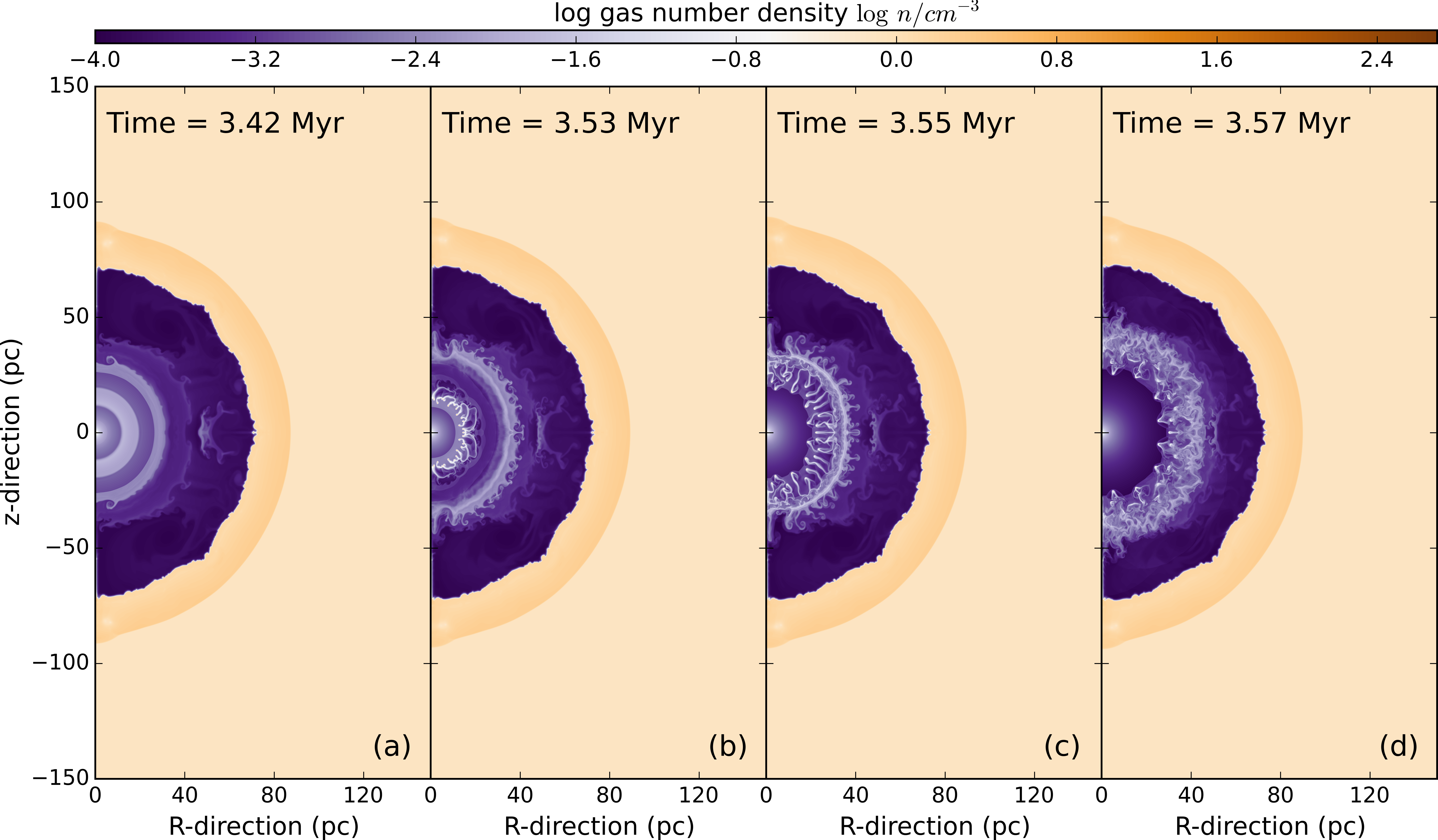}
        \caption{Time sequence of density \mpo{maps} (in $\rm cm^{-3}$) \mpo{with} circumstellar shells released inside the stellar wind bubble 
        	 generated by our massive star at rest ($v_{\star}=0\, \rm km\, \rm s^{-1}$), while \mpo{it passes through} two successive luminous-blue-variable phases before the Wolf-Rayet wind blows. 
		 The panels are shown at times $3.42$ (a), $3.53$ (b), $3.55$ (c) and $3.57\, \rm Myr$ (d), respectively. 
		 \textcolor{black}{On each panel the star is located at the origin. }
                 }      
        \label{fig:presn2}  
\end{figure*}


\section{Method}
\label{sect:method}

This section introduces the reader to the methods used to perform hydrodynamical simulations of the 
surroundings of a massive star undergoing both luminous blue variable and Wolf-Rayet phases. The evolution of the 
circumstellar medium is simulated from the zero-age main sequence of the progenitor to the \mpo{late phase of supernova 
remnant evolution, for several proper-motion speeds} of the driving star.

\subsection{Simulation method for the pre-supernova phase}
\label{sect:hydro}

The structure of the pre-supernova circumstellar medium arises from stellar wind-ISM interaction. \mpo{We model it with the same method as was used} for studying bow shocks of runaway massive stars, summarised in~\citet{meyer_obs_2016}. 
The stellar surroundings are simulated with axisymmetric, two-dimensional numerical hydrodynamics simulations 
performed with the {\sc pluto}\footnote{http://plutocode.ph.unito.it/} code~\citep{mignone_apj_170_2007, migmone_apjs_198_2012}. 
We use \textcolor{black}{cylindrical coordinates with a uniform grid} \textcolor{black}{$[O;R_{\rm max}]\times[z_{\rm min};z_{\rm 
max}]$ that is discretised with $N_{\rm R}\, \times\, N_{\rm z}$} grid zones. The uniform spatial mesh 
resolution is $\Delta=R_{\rm max}/N_{\rm R}=(z_{\rm max}-z_{\rm min})/N_{\rm z}$. 
The supersonic stellar wind is injected into the computational domain via a \mpo{sphere of $20$ cells in radius centered 
at} the origin of the grid and filled with the wind density profile, 
\begin{equation}
	\rho_{w} = \frac{ \dot{M} }{ 4\pi r^{2} v_{\rm w} },
\label{eq:wind}
\end{equation}
where $\dot{M}$ denotes the time-dependent mass-loss rate of the massive star, $r$ is the distance to the origin of the domain, $O$, 
and $v_{\rm w}$ is the wind velocity. \mpo{The same} method has been used in~\citet{comeron_aa_338_1998} and~\citet{vanmarle_aa_469_2007,vanmarle_apj_734_2011, 
vanmarle_aa_561_2014}. 
The stellar wind parameters $\rho_{w}$ and $v_{\rm w}$ are time-dependently interpolated from the tabulated stellar evolution model 
of a non-rotating $60\, \rm M_{\odot}$ star presented in~\citet{groh_aa564_2014}.
The wind-blown bubble resulting from the wind-ISM interaction is computed in the \mpo{rest} frame of the central star, 
and its proper stellar motion $v_{\star}$ is taken into account by setting an inflowing medium at the 
$z=z_{\rm max}$ boundary. \textcolor{black}{Outflow} boundary conditions are assigned at $z=z_{\rm min}$ and $R=R_{\rm max}$, respectively. 
The stellar wind material injected into the domain is distinguished from the ambient medium using a passive 
tracer, \mpo{$Q_{1}$, that is initially set to $Q_{1}(\bmath{r})=1$ in the wind material and 
to $Q_{1}(\bmath{r})=0$ in the ISM gas, respectively. We follow the mixing of materials in the circumstellar structure using the advection equation,}
\begin{equation}
	\frac{\partial (\rho Q_{1}) }{\partial t } +  \bmath{ \nabla } \cdot  ( \bmath{v} \rho Q_{1}) = 0.
\label{eq:tracer}
\end{equation}
\textcolor{black}{
The computationally-intensive calculations were performed on the HPC systems Cobra using Intel Skylake processors 
and Draco using Intel Haswell, Broadwell processors at the Max Planck Computing and Data Facility 
(MPCDF\footnote{https://www.mpcdf.mpg.de/}) in Garching,   
and on the North-German Supercomputing  Alliance (HLRN\footnote{https://www.hlrn.de/}) using the HPC compute system 
in Berlin operating with Cray XC40/30 processors, respectively.  
}

\subsection{Setting up the supernova explosion}
\label{subsect:sn}

\dmam{
Once the pre-supernova circumstellar medium has been modelled, we \mpon{insert} a spherically-symmetric explosion into a spherically-symmetric stellar wind in order to separately calculate in 1D fashion the initial expansion of the supernova shock wave into the progenitor's last freely-expanding wind, whose solution is used as initial condition for the 2D supernova remnants~\citep{meyer_mnras_450_2015}. 
}
The expanding blastwave is characterized by its energy, taken to be $E_{\rm ej}=10^{51}\, \rm erg$,
and by the mass of the ejecta released at the time of the explosion, $M_{\rm ej}$,  estimated as
\begin{equation}
   M_{\rm ej} =  M_{\star} - \int_{t_\mathrm{ZAMS}}^{t_\mathrm{SN}} \dot{M}(t)~ dt - M_{\mathrm{NS}} = 11.1\, \rm M_{\odot},
   \label{eq:co}
\end{equation}
where $t_\mathrm{ZAMS}$ and $t_\mathrm{SN}$ are the zero-age-main-sequence and supernova times, respectively, \dmam{and 
where $M_\mathrm{NS}=1.4\, \rm M_{\odot}$ is the mass of the remnant neutron star}.

The expansion of the supernova shock wave is \mpo{launched} using the method of~\citet{whalen_apj_682_2008} and~\citet{vanveelen_aa_50_2009}, in which a blastwave density profile, $\rho(r)$, \dmam{is deposited on the 
top of the progenitor's wind profile taken from the modelled nebulae, at the pre-supernova time. We ensure that the outer boundary of this 
1D computational domain is smaller than the value of the wind termination shock of the pre-shaped 
circumstellar medium. 
The ejecta profile is defined by two characteric lengths corresponding to the radius of the progenitor's 
core at the moment of the explosion, 
$r_\mathrm{core}$, and the position 
of the forward shock of the blastwave, $r_\mathrm{max}$. 
The simulation is initialised at time $t_{\rm max}=r_{\rm max}/v_{\rm max}$, with $v_{\rm max}=30000\, \rm km\, \rm s^{-1}$ the ejecta \mpon{top speed}~\citep{vanveelen_aa_50_2009}. 
The quantity $r_{\rm max}$ is a free parameter determined according to the numerical procedure described in~\citet{whalen_apj_682_2008}.  
}
The density profile of the ejecta \mpo{follows a} piece-wise function, 
\begin{equation}
\rho(r) = \begin{cases}
        \rho_{\rm core}(r) & \text{if $r \le r_{\rm core}$ },               \\
        \rho_{\rm max}(r)  & \text{if $r_{\rm core} < r < r_{\rm max}$},    \\
        \rho_{\rm csm}(r)  & \text{if $r \ge r_{\rm max}$},                 \\
        \end{cases}
	\label{cases}
\end{equation}
with \mpo{a constant density,
\begin{equation}
   \rho_{\rm core}(r) =  \frac{1}{ 4 \pi n } \frac{ (10 E_{\rm ej}^{n-5})^{-3/2}
 }{  (3 M_{\rm ej}^{n-3})^{-5/2}  } \frac{ 1}{t_{\rm max}^{3} },
   \label{sn:density_1}
\end{equation}
imposed for the plateau of the ejecta} in the $[O;r_{\rm core}]$ region of the domain. The quantity 
\begin{equation}
   \rho_{\rm max}(r) =  \frac{1}{ 4 \pi n } \frac{ (10 E_{\rm
ej}^{n-5})^{(n-3)/2}  }{  (3 M_{\rm ej}^{n-3})^{(n-5)/2}  } 
\bigg(\frac{r}{t_{\rm max}}\bigg)^{-n},
   \label{sn:density_2}
\end{equation}
is a function of $r_{\rm core}$ and $r_{\rm max}$~\citep{truelove_apjs_120_1999}, where 
the index, $n$, is set to $n=11$ that is typical for core-collapse supernova explosions~\citep{chevalier_apj_258_1982}. 
In Eqs.~(\ref{cases}), $\rho_{\rm cms}$ is the freely-expanding wind profile measured from the pre-supernova simulations 
of the circumstellar medium. 
The ejecta \mpo{speed follows a homologous-expansion} profile,  
\begin{equation}
   v(r) = \frac{ r }{ t },
   \label{sn:vel}
\end{equation}
$\mathrm{for}\,  \mathrm{times}\, t>t_{\rm max}$, where $t$ is the time after the supernova explosion. 
The velocity of the ejecta at the location of $r_{\rm core}$ reads,
\begin{equation}
   v_{\rm core} = \bigg(  \frac{ 10(n-5)E_{\rm ej} }{ 3(n-3)M_{\rm ej} } 
\bigg)^{1/2},
   \label{sn:vcore}
\end{equation}
see~\citet{truelove_apjs_120_1999}. 
\textcolor{black}{
The 1D solution for the ejecta-wind interaction is mapped onto the 2D domain when the forward shock of the 
expanding supernova blastwave reaches $8\, \rm pc$. With the grid resolution that we consider, the 
mapped ejecta profiles are hence resolved with $\sim 160$ grid zones. 
}

Last, a second passive scalar $Q_{2}(\bmath{r})$ is introduced, obeying the advection equation
\begin{equation}
	\frac{\partial (\rho Q_{2}) }{\partial t } +  \bmath{ \nabla } \cdot  ( \bmath{v} \rho Q_{2}) = 0,
\label{eq:tracers}
\end{equation}
which distinguishes the ejecta from the stellar wind and the ambient medium materials. It is initially set to 
$Q_{2}(\bmath{r})=1$ for the region made of supernova ejecta material and to $Q_{2}(\bmath{r})=0$ in the other part of the 
computational domain, respectively.

\begin{figure}
        \centering
                \includegraphics[width=0.899\columnwidth]{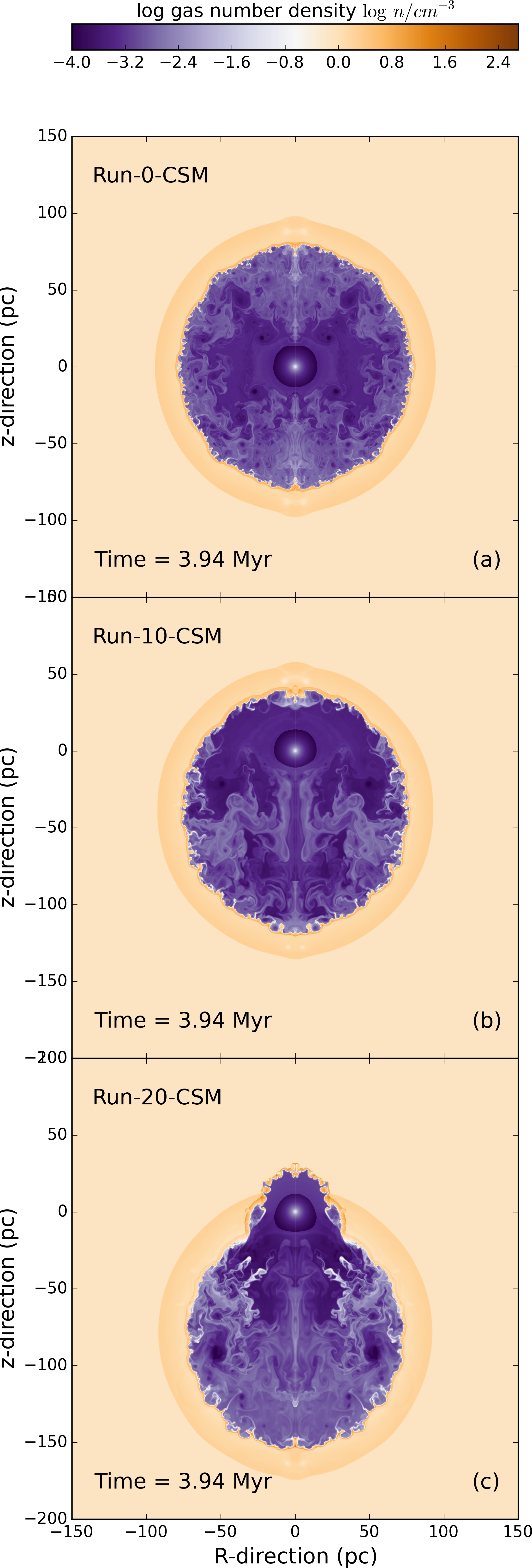}
        \caption{Density field (in $\rm cm^{-3}$) of the circumstellar medium of a massive star moving with 
        	 $v_{\star}=0$ (a), $10$ (b), and $20\, \rm km\, \rm s^{-1}$ (c) at \dmam{the pre-supernova time}.
        	 \textcolor{black}{On each panel the star is located at the origin.}}
        \label{fig:presn3}  
\end{figure}

\subsection{Governing equations}
\label{subsect:goveq}

We model the stellar wind-ISM interaction and the supernova remnant within the framework of the non-ideal hydrodynamics, i.e. by 
solving the Euler equations and by accounting for energy losses by optically-thin radiative cooling. We 
solve the set of equations composed of the equation for the conservation of mass, 
\begin{equation}
	   \frac{\partial \rho}{\partial t}  + 
	   \bmath{\nabla}  \cdot (\rho\bmath{v}) =   0,
\label{eq:euler1}
\end{equation}
of linear momentum,
\begin{equation}
	   \frac{\partial \rho \bmath{v} }{\partial t}  + 
           \bmath{\nabla} \cdot ( \bmath{v} \otimes \rho \bmath{v}) 	      + 
           \bmath{\nabla}p 			      =   \bmath{0},
\label{eq:euler2}
\end{equation}
and of the total energy,
\begin{equation}
	  \frac{\partial E }{\partial t}   + 
	  \bmath{\nabla} \cdot(E\bmath{v})   +
	  \bmath{\nabla} \cdot (p \bmath{v})   =	   
	  \itl{\Phi}(T,\rho),
\label{eq:euler3}
\end{equation}
where
\begin{equation}
	E = \frac{p}{(\gamma - 1)} + \frac{\rho v^{2}}{2},
\label{eq:energy}
\end{equation}
and $\rho$ is the mass density of the gas, $p$ its pressure, 
$\bmath{v}$ the vector velocity, respectively. The gas temperature reads,
\begin{equation}
	T =  \mu \frac{ m_{\mathrm{H}} }{ k_{\rm{B}} } \frac{p}{\rho},
\label{eq:temperature}
\end{equation}
where $k_{\rm B}$ is the Boltzmann constant and $\mu$ is the mean molecular 
weight, so that the gas mass density reads,   
\begin{equation}
	\rho= \mu \, n\, m_{\rm H},
\label{eq:mudef}
\end{equation}
with $n$ the total number density of the plasma and $m_{\rm H}$ the mass of a hydrogen atom. 
The adiabatic index of the gas is set to $\gamma=5/3$.  
%
%
%
As in~\citet{meyer_mnras_450_2015}, we use a finite-volume method with the Harten-Lax-van Leer approximate Riemann solver, 
and integrate the Euler equations with a second order, unsplit, time-marching algorithm.

Our simulations take into account internal energy losses and gains by optically-thin 
cooling and heating, respectively. These mechanisms are represented with the right-hand 
term of Eq.~(\ref{eq:euler3}), 
\begin{equation}  
	 \itl \Phi(T,\rho)  =  n_{\mathrm{H}}\itl{\Gamma}(T)   
		   		 -  n^{2}_{\mathrm{H}}\itl{\Lambda}(T),
\label{eq:dissipation}
\end{equation}
where $\itl{\Gamma}(T)$ and $\itl{\Lambda}(T)$ are the heating and cooling components of the expression, respectively, 
and where $n_{\mathrm{H}}$ is the hydrogen number density.
The contribution for heating, $\it \Gamma$, mimics the reionisation of recombining hydrogen atoms by 
the photon field of the hot star~\citep{osterbrock_1989,hummer_mnras_268_1994}. 
The cooling term is based on the prescriptions by optically-thin cooling derived by~\citet{wiersma_mnras_393_2009},  
accounting for the radiation losses for hydrogen and helium at $T<10^{6}\, \rm K$ and for metals at 
$T \ge 10^{6}\, \rm K$~\citep{wiersma_mnras_393_2009}.  
The cooling curve has been updated with collisionally excited forbidden lines such as [O{\sc iii}] $\lambda \, 5007$ line 
emission~\citep{henney_mnras_398_2009}.  
  
\begin{figure}
        \centering
                \includegraphics[width=1.0\columnwidth]{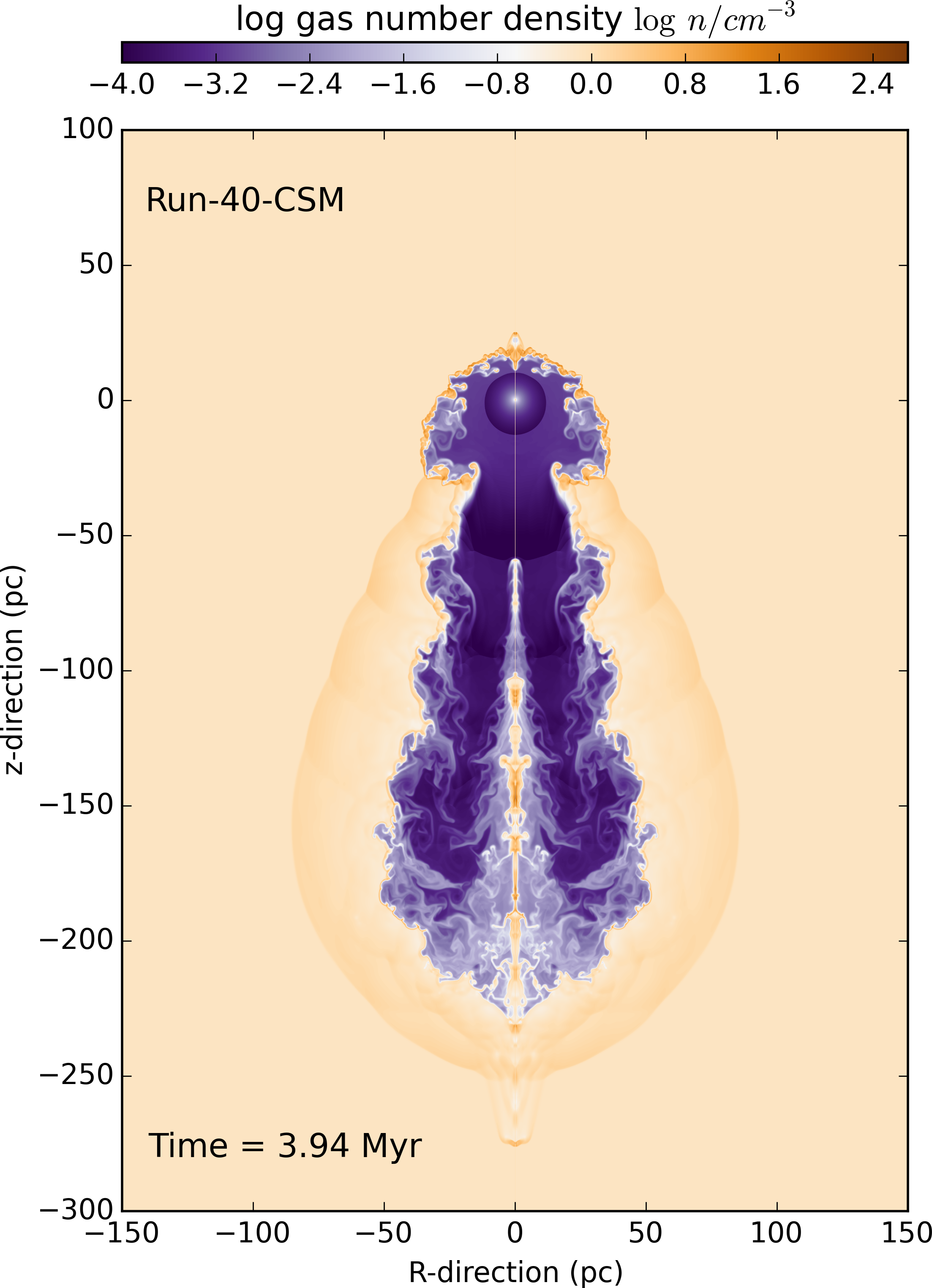}
        \caption{
		 Same as Fig.~\ref{fig:presn3} for a star moving with $v_{\star}=40\, \rm km\, \rm s^{-1}$. 
		 \textcolor{black}{The star is located at the origin. }
                 }      
        \label{fig:presn4}  
\end{figure}

\subsection{Simulation parameters}
\label{sect:para}

Fig.~\ref{fig:star} plots the stellar surface properties that are used as boundary conditions in our hydrodynamical 
simulations. The star's post-main-sequence mass history (panel a, in $\rm M_{\odot}$), mass-loss rate evolution (panel b, in $\rm M_{\odot}\, \rm yr^{-1}$), 
and wind velocity (panel c, in $\rm km\, \rm s^{-1}$) are plotted as a function of time (in $\rm Myr$) starting 
from $3\, \rm Myr$. Earlier times correspond to the O-type main-sequence phase of the star during which the stellar wind properties do not evolve much. 
The red dots on the figures denote the principal evolutionary phases and the yellow labels on the figures 
indicate the corresponding spectral type. 
The stellar evolution model of this zero-age-main-sequence non-rotating $60\, \rm M_{\odot}$ star has been computed up to 
the Si burning phase with the {\sc genec} code~\citep{ekstroem_aa_537_2012}. The effective temperature and the terminal wind velocity 
have been estimated using atmospheric radiative transfer modelling and prescriptions derived from observations of stellar 
populations~\citep{groh_aa564_2014}. 
\mpo{They are} characterised by a long main-sequence phase lasting \mpo{approximately}
$3\, \rm Myr$ during which the star blows winds of $\dot{M}\approx 10^{-6}$$-$$10^{-5.5}\, \rm M_{\odot}\, \rm yr^{-1}$, 
$v_{\rm w}\approx 3500$-$2000\, \rm km\, \rm s^{-1}$ and loses roughly $8\, \rm M_{\odot}$ of material. 
Beginning of the last $\rm Myr$, the mass-loss increases to $\dot{M}\approx 10^{-5}\, \rm M_{\odot}\, \rm yr^{-1}$, 
while $v_{\rm w}\approx 1300\, \rm km\, \rm s^{-1}$, and the star adopts a B spectral type. This is followed by two consecutive 
luminous-blue-variable eruptions with $\dot{M}\approx 10^{-3.5}$-$10^{-3}\, \rm M_{\odot}\, \rm yr^{-1}$ with wind velocity 
decreasing down to \mpo{nearly} $ 200\, \rm km\, \rm s^{-1}$. 
The star then evolves to the so-called Wolf-Rayet phase of both high mass-loss rate and a large wind velocity. 
\dmam{
We do not consider the last pre-supernova increase of the stellar wind velocity up to $\approx 5000\, \rm km\, s^{-1}$ in 
this study, as very few examples of such strong-winded Wolf-Rayet stars have been so far monitored~\citep{2000A&A...360..227N,2015A&A...578A..66T}.
}

The free parameter in our series of 2D computationally-intensive simulations is the space velocity of the star, $v_{\star}$, which spans from 
the static case producing a spherical wind-blown bubble ($v_{\star}=0\, \rm km\, \rm s^{-1}$) to the runaway case generating an bow shock 
($v_{\star}=40\, \rm km\, \rm s^{-1}$). 
In all our models, as in the series of papers devoted to a grid of bow shocks models of~\citet{meyer_2014bb,meyer_obs_2016}, the 
massive star is assumed to be located in the warm phase of the ISM, characterised by a temperature of $ 8000\, \rm K$ and a 
number density of $0.79\, \rm cm^{-3}$. 
After the modelling of the circumstellar medium of the evolving massive star up the pre-supernova phase, the solution 
is mapped onto a larger computational domain that is filled with unperturbed ambient ISM material. The supernova explosion 
is setup according to the method developed in~\citet{meyer_mnras_450_2015} and described in the paragraphs above. 
Our simulation runs and their characteristics are summarised in Table~\ref{tab:models}.

\section{Results}
\label{sect:snr}

In this section, we describe the pre-supernova circumstellar evolution of our $60\, \rm M_{\odot}$ star 
and how the expanding supernova blast wave interacts with it. We present the thermal X-ray signatures of 
the supernova remnants and analyse the mixing of stellar wind, ISM material, and supernova ejecta.

%

%
%


\subsection{The pre-supernova circumstellar medium}
\label{sect:cms}

Our model Run-0-CSM for a static massive star develops a spherical wind bubble of swept-up wind and ISM gas. 
As the wind properties do not evolve much during the main-sequence phase~\citep{groh_aa564_2014}, 
the bubble nebula grows according to the picture described in~\citet{weaver_apj_218_1977}. 
When the OB-type star further evolves, the successive changes in the stellar properties produce several shells that expand 
at different velocities and eventually collide. Fig.~\ref{fig:presn2} presents the bubble density field of Run-0-CSM for different 
time instances: at time $3.42\, \rm Myr$ the main-sequence bubble is constituted of an outer dense cold shell of radius \mpo{around}
$60\, \rm pc$ surrounding a hot diluted region at radii $25$-$50\, \rm pc$, inside of which the post-main-sequence 
shells are released (Fig.~\ref{fig:presn2}a). 
The dense \mpo{shells collide with each other (Fig.~\ref{fig:presn2}b) and then develop instabilities (Fig.~\ref{fig:presn2}c,d) as} described in~\citet{garciasegura_1996_aa_305,vanmarle_aa_469_2007,2012A&A...547A...3V}. At the 
pre-supernova time, the circumstellar medium has adopted the shape of a spherically-symmetric structure in which the expanding 
shells have reached the contact discontinuity of the main-sequence bubble separating hot shocked wind and cold shocked ISM (Fig.~\ref{fig:presn3}a). 
The slow stellar motion of $10\, \rm km\, \rm s^{-1}$ in Run-10-CSM is \dmam{not} sufficient to break this spherical symmetry 
(Fig.~\ref{fig:presn3}b). When the star moves into its own wind bubble, the post-main-sequence shells are released 
from \mpo{an off-center} location in the bubble. 
The shell of colliding winds interact first with the bubble in the direction of motion of the star. This rapidly \mpo{destabilises the 
contact discontinuity of the nebula ahead of the star's direction of motion, whereas in the opposite direction} the post-main-sequence material 
interacts later, as a natural consequence of \dmam{the off-centered position of the star in the nebula} (Fig.~\ref{fig:presn3}b).

\begin{figure*}
        \centering
        \begin{minipage}[b]{ 1.0\textwidth}  \centering
                \includegraphics[width=1.0\textwidth]{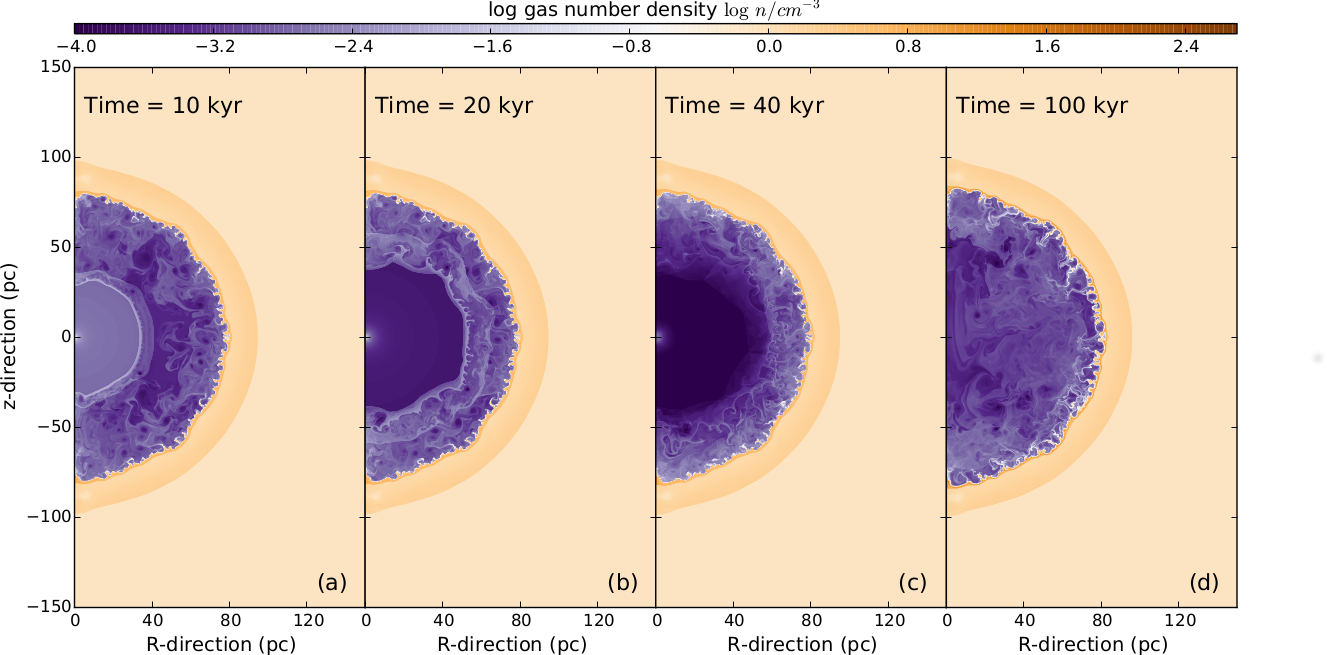}
        \end{minipage} \\         
        \caption{ 
        	 Density fields (in $\rm cm^{-3}$) in the supernova remnant of a $60\, \rm M_{\odot}$ star \mpo{at rest throughout its entire pre-supernova evolution. Panel a) (left) shows the reflection of the shock wave off the perimeter of the wind bubble at time $10\, \rm kyr$,  and panel b) (right) displays the endstate}
        	 after \dmam{internal reflection at time $100\, \rm kyr$}. }      
        \label{fig:snr_0kms}  
\end{figure*}

Fig.~\ref{fig:presn3}c reports the pre-supernova circumstellar medium in our model \mpo{Run-20-CSM} in which the star moves through the ISM 
with Mach number ${\it M} =v_{\star}/c_{\rm s}=1$. The main-sequence stellar-wind bubble is still spherical, and of size similar to  
that in Run-10-CSM. The star has traveled $3\, \mathrm{Myr}\, \times 20\,  \mathrm{km}\, \mathrm{s}^{-1} \simeq 60 \, \rm pc$ 
during its main-sequence phase, which is \mpo{approximately} the final radius of its own wind bubble. The \mpo{shells of 
luminous-blue-variable and Wolf-Rayet material are hence released} directly in the dense cold shocked ISM gas, not in the stellar 
wind cavity (Fig.~\ref{fig:presn3}a,b). 
The shells expand freely in the unperturbed ISM ahead of the direction of stellar motion and develop Rayleigh-Taylor instabilities, whereas 
they fill the interior of the main-sequence wind bubble in the opposite direction. The overall shape of the circumstellar medium therefore becomes 
strongly asymmetric, i.e. it is shaped as a bubble of cold OB gas \mpo{with embedded} enriched post-main-sequence material, ahead of which 
a bow shock of \dmam{thin, instable but dense material forms into the ambient medium, as a consequence of the high Wolf-Rayet wind density 
($\dot{M}\approx 10^{-5}\, \rm M_{\odot}\, \rm yr^{-1}$)}. 
When the star moves even faster, e.g. with velocity $v_{\star}=40\, \rm km\, \rm s^{-1}$, the luminous blue variable and Wolf-Rayet shells are expelled out of the main-sequence nebula and expand quasi-spherically into the ISM, see model \mpo{Run-40-CSM} (Fig.~\ref{fig:presn4}). The pre-supernova circumstellar medium is completely asymmetric, and only a narrow region of the post-main-sequence ring is connected to the drop-like nebula that reflects the distortion of the wind bubble by the stellar motion. 
\textcolor{black}{
Note that a boundary effect develops as material accumulates close to the symmetry axis (Figs.~\ref{fig:presn3},~\ref{fig:presn4}). 
This well-known problem~\citep{mackey_sept_2014} is unavoidable in two-dimensional axisymmetric simulations of this kind using uniform grids. It can be circumvented by means of three-dimensional calculations, however at the cost of losses in spatial resolution~\citep{2019A&A...631A.170R}. 
}

\begin{figure*}
        \centering
                \includegraphics[width=0.925\textwidth]{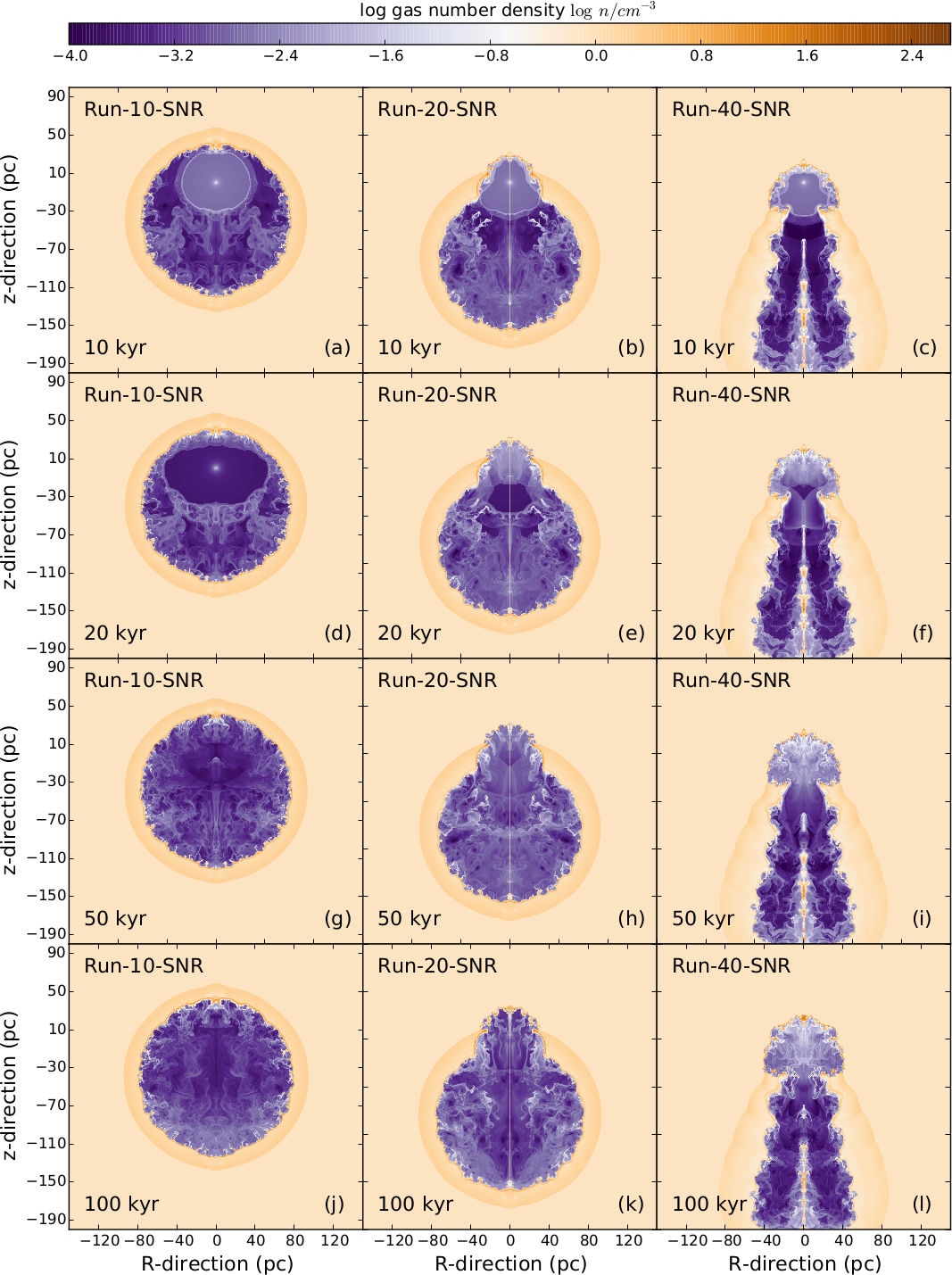}
        \caption{Density fields (in $\rm cm^{-3}$) of the supernova remnants generated by a moving $60\,$-$ \rm M_{\odot}$ stars. 
        	 The columns display the models Run-10-SNR (left, $v_{\star}=10\, \rm km\, \rm s^{-1}$), 
        	 model Run-20-SNR (middle, $v_{\star}=20\, \rm km\, \rm s^{-1}$) and model Run-40-SNR (right, $v_{\star}=40\, \rm km\, \rm s^{-1}$). 
        	 Each remnant is shown at times \textcolor{black}{$10\, \rm kyr$ (top row, panels a-c), $20\, \rm kyr$ (second row, panels d-f), 
        	 $50\, \rm kyr$ (third row, panels g-i), and $100\, \rm kyr$ (bottom row, panels j-l)}, respectively. 
                 }      
        \label{fig:snr}  
\end{figure*}

\subsection{Dynamics of asymmetric supernova remnants}
\label{sect:snr_dyn}

If one neglects asymmetries potentially developing in the supernova explosion itself~\citep{toledo_mnras_442_2014}, 
the remnant of a massive progenitor star at rest conserves its sphericity as the forward shock wave expands 
symmetrically into the wind cavity. 
At time $10\, \rm kyr$ \mpo{it interacts with the dense region of the progenitor's wind bubble (Fig.~\ref{fig:snr_0kms}a) 
and is reverberated towards the center of the explosion. Subsequent multiple internal reflections of the shock wave generate
a large turbulent} region (Fig.~\ref{fig:snr_0kms}b). 
In Fig.~\ref{fig:snr} we show time sequences of the density field (rows) of a supernova remnant produced by a 
$60$-$ \rm M_{\odot}$ star moving with different space velocities (columns). The bulk motion of the star spans 
from $v_{\star}=10$ (left column) to $40\, \rm km\, \rm s^{-1}$ (right column), and the remnants are displayed 
from $10\, \rm kyr$ (top row) up to $100\, \rm kyr$ (bottom row) after the explosion.

Our model Run-10-SNR with a slowly-moving star is plotted in the first column of panels of Fig.~\ref{fig:snr}. 
The shock wave is clearly identifiable in Fig.~\ref{fig:snr}a as it has not yet interacted with the termination 
shock of the wind bubble. \mpo{Reflected} shocks forms when the shock wave collides with termination shock of 
the wind bubble and with the dense main-sequence shell, respectively, making interior of the bubble asymmetric 
(Fig.~\ref{fig:snr}d). 
At later times ($50$-$100\, \rm kyr$), the part of the shock wave \dmam{which propagates along the stellar 
direction of motion through the bubble, penetrates the dense layer of shock ISM gas and makes it slightly denser as it 
had more time to interact with it, while the interior of the remnant is filled with a mixture of low-density, 
high-temperature ejecta and \textcolor{black}{filamentary} shocked wind material which keeps on melting with each other} (Fig.~\ref{fig:snr}g,j).

\begin{figure*}
        \centering
                \includegraphics[width=0.8\textwidth]{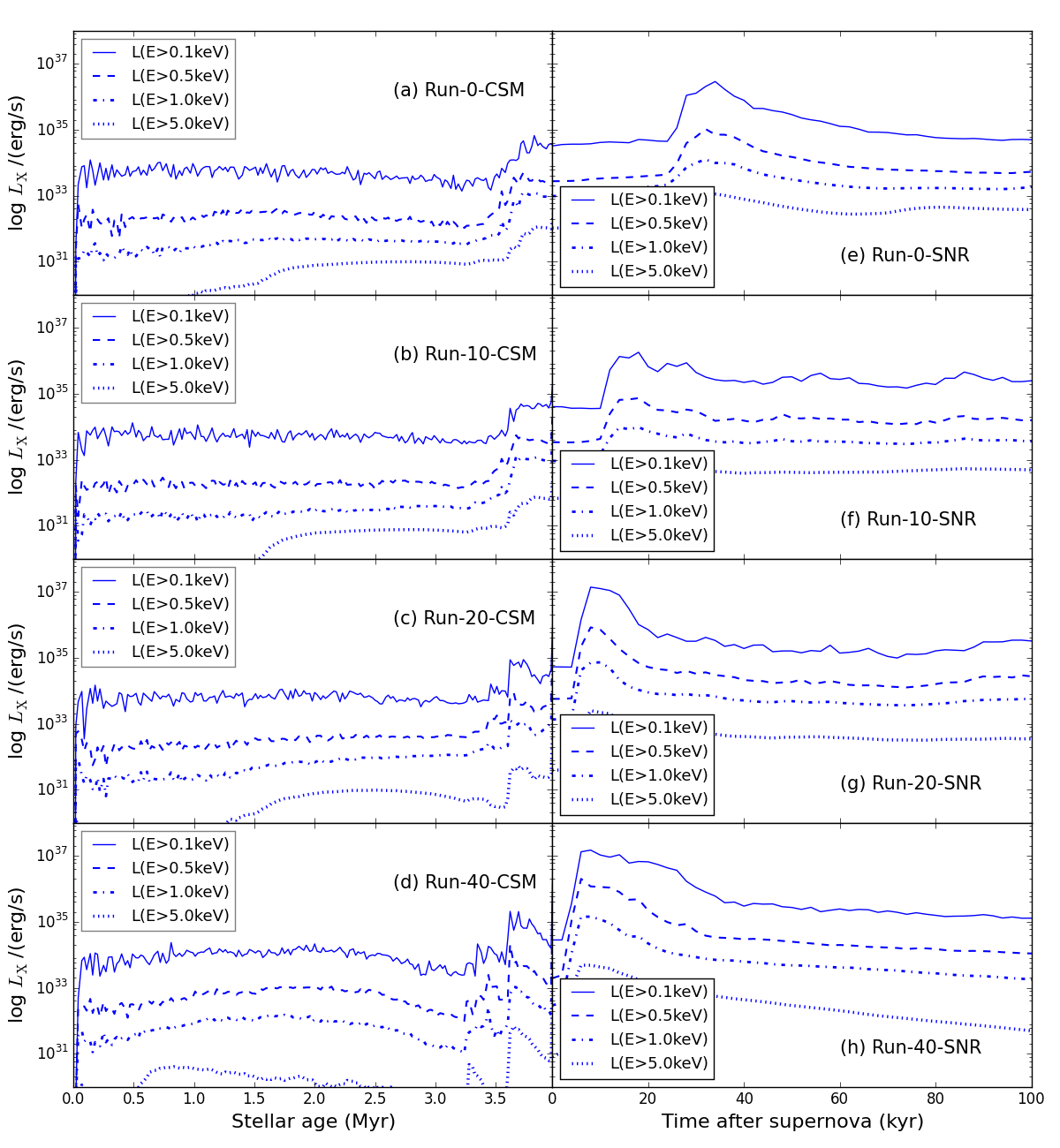}
        \caption{\mpo{Lightcurves of thermal X-ray before (left panels) and after the supernova explosion (right panels). 
        	 The integrated luminosity is} plotted as a function of time, for different energy bands (in $\rm erg\, \rm s^{-1}$, blue lines) and 
        	 for several space velocities of the progenitor, spanning from $v_{\star}=0\, \rm km\, \rm s^{-1}$ (top panels) 
        	 to $v_{\star}=40\, \rm km\, \rm s^{-1}$ (bottom panels), respectively. 
                 }      
        \label{fig:luminosities}  
\end{figure*}

In our model Run-20-SNR the ejecta first fill the protuberance generated by the \dmam{evolved} Wolf-Rayet wind (Fig.~\ref{fig:snr}b). \mpo{As for a slower} supernova progenitor, the shock wave progressively penetrates and expands into the entire upper cavity of the remnant which turns into a hot region of mixed wind and ISM material. 
The shock wave \mpo{moves} back and forth inside the post-main-sequence mushroom as it experiences multiple reflections 
between the center of the explosion and the walls of the protuberance (Fig.~\ref{fig:snr}e,h). Once it goes 
through, a transmitted shock front freely expands into the unperturbed ISM and \dmam{begins to locally} recovers a global 
spherical aspect, albeit affected by \textcolor{black}{Rayleigh-Taylor} instabilities (Fig.~\ref{fig:snr}k).  
Simultaneously, the unstable shock wave expands downstream the progenitor star's direction inside the cavity 
\dmam{and eventually reaches the bottom of the wind bubble (Fig.~\ref{fig:snr}k)}. 

In our fast-moving model Run-40-SNR, the supernova shock first fills the Wolf-Rayet bubble, which gets 
heated by reverberations against the unstable post-main-sequence ring (Fig.~\ref{fig:snr}c,f,i). It is then channeled 
as a jet-like extension to the spherical region of shocked ejecta into the tubular cavity formed by the 
tail of the progenitor's bow shock with the mechanism described in~\citet{blandford_301_natur_1983,cox_mnras_250_1991}. 
Last, the reflection of the shock wave that also happens inside the tail of the bow shock 
fills the main-sequence nebula with hot gas as \mpo{described} in~\citet{meyer_mnras_450_2015}, see Fig.~\ref{fig:snr}l.

\begin{figure*}
        \centering
                \includegraphics[width=0.8\textwidth]{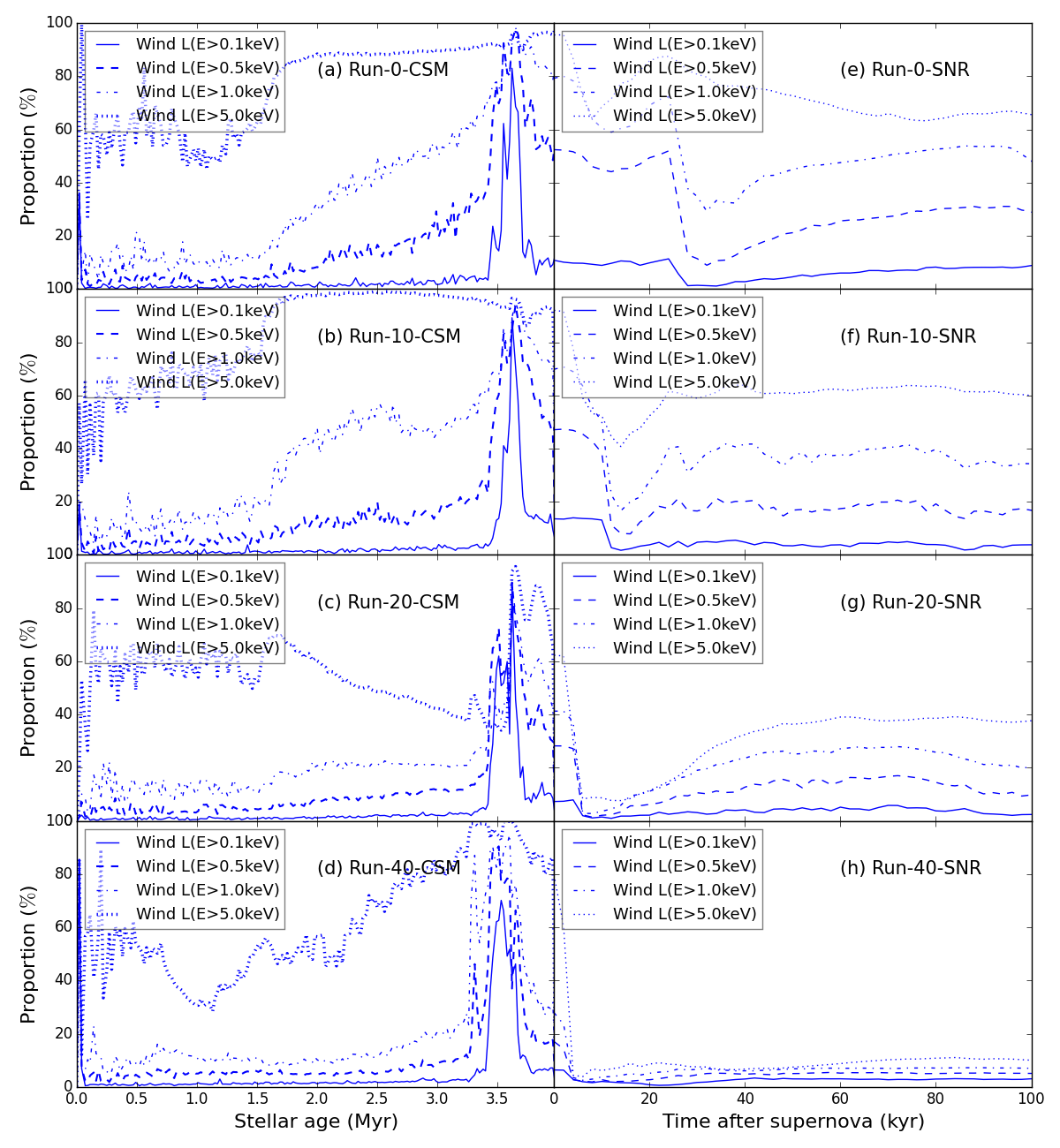}
        \caption{Stellar wind contributions to the thermal X-ray luminosity \mpo{(in $\%$) as a function of time for the pre-supernova epoch (left panels) 
		 and the supernova remnant phase (right panels). 
        Each panel covers several energy bands, and each row shows} results for different space velocities of 
        	 the progenitor, spanning from $v_{\star}=0\, \rm km\, \rm s^{-1}$ (top panels) to $v_{\star}=40\, \rm km\, \rm s^{-1}$ (bottom panels), respectively. }      
        \label{fig:luminosities_tracers}  
\end{figure*}

\subsection{X-ray signature}
\label{sect:snr_xr}

The thermal X-ray luminosity for the circumstellar nebulae and the supernova remnants of our massive star 
is plotted as a function of time in Fig.~\ref{fig:luminosities}. 
To predict the emission properties of our old ($10-100\, \rm kyr$) supernova remnants, we generate X-ray lightcurves \citep[cf.][]{meyer_mnras_450_2015}. For each simulation snapshot, the X-ray \mpo{emission coefficient, \textcolor{black}{$j_{\mathrm{X}}^{\mathrm{E}\ge\alpha}$} for the \textcolor{black}{$\mathrm{E}\ge\alpha$} 
energy band,} is calculated and integrated over the whole nebula, 
\begin{equation}
        L_{X}^{\textcolor{black}{\mathrm{E}\ge\alpha}} =  \iint_{\mathrm{SNR}} j_{\mathrm{X}}^{\textcolor{black}{\mathrm{E}\ge\alpha}}(T) n_{\mathrm{H}}^{2}  \mathrm{dV},
        \label{eq:lum_x}  
\end{equation}
where $n_{\rm H}$ is the hydrogen number density in the supernova remnant (SNR). 
The thermal X-ray emission coefficient of the diffuse ISM is tabulated as a function of temperature with the 
{\sc xspec}\footnote{https://heasarc.gsfc.nasa.gov/xanadu/xspec/} software~\citep{arnaud_aspc_101_1996} which uses the
solar abundances of~\citet{asplund_araa_47_2009}.

\begin{figure*}
        \centering
        \includegraphics[width=0.875\textwidth]{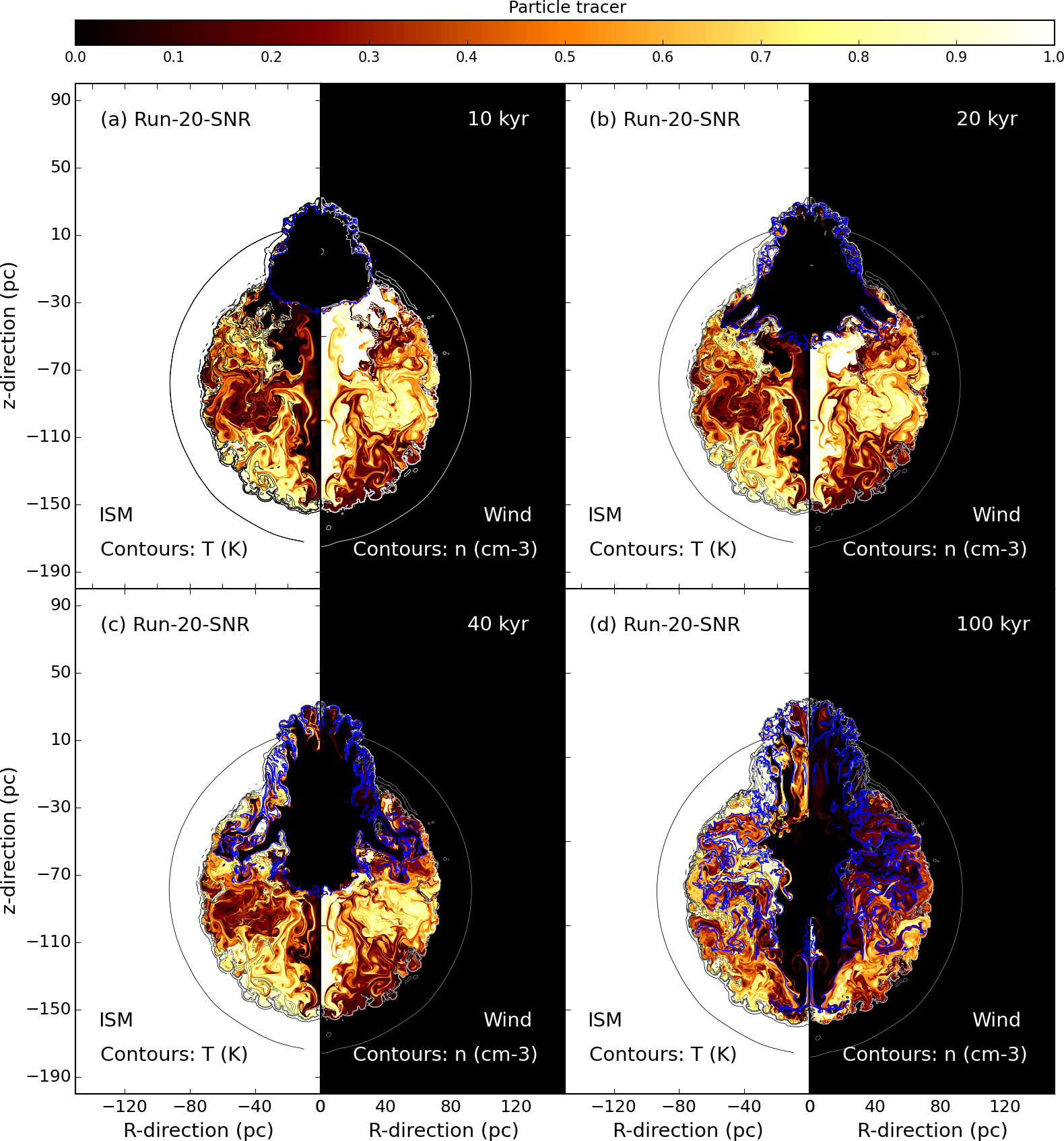}
        \caption{Mixing of material in the supernova remnant generated by a progenitor moving with velocity $20\, \rm km\, \rm s^{-1}$, shown at times $10\, \rm kyr$ (a), \textcolor{black}{$20\, \rm kyr$} (b), $50\, \rm kyr$ (c) and $100\, \rm kyr$ (d). 
        	 The figures plot the value of the quantity \textcolor{black}{1-$Q_{1}$ representing the proportion of ISM material (left part of the panels) and the value of the passive scalar tracer  $Q_{1}$ standing for the proportion of stellar wind gas (right part of the panels)}, respectively. 
        	 The overplotted black contours indicate the temperature ($T=10^{5}$, $10^{6}$, $10^{7}\, \rm K$, left part of the panels) 
        	 while the white contours stands for the gas number density ($n=1.0$, $10^{1}$, $10^{2}$ and $10^{3}\, \rm cm^{-3}$, right part of the panels). 
        	 The blue contours \textcolor{black}{of the passive scalar field $Q_{2}$} \mpo{indicate a $10\%$ contribution} of ejecta in number density. 
                 }      
        \label{fig:snr_20kms_mixing}  
\end{figure*}

The pre-supernova evolution of the thermal X-ray luminosity in our model Run-0-CSM, in which the star is at rest, begins by a rapid 
increase up to $L_{\rm X}^{\mathrm{E\ge0.1\, \rm keV}}\approx 10^{34}\, \rm erg\, \rm s^{-1}$ \mpo{during the initial expansion of the wind bubble, followed by a nearly constant brightness} 
throughout the entire main-sequence phase of the star (up to $\approx 3.2\, \rm Myr$). The small \mpo{luminosity variations during this phase} are caused by the instabilities in the shocked ISM. When the star moves quickly, \mpo{an important mechanism is} vortex 
shedding, described in~\citet{wareing_apj_660_2007} and~\citet{2019A&A...625A...4G}.  
A slight increase of the luminosity $L_{X}^{\mathrm{E}\ge0.1\mathrm{keV}}$ \mpo{coincides} with the post-main-sequence evolutionary phases of the star, when the nebula reaches 
$L_{X}^{\mathrm{E}\ge0.1\mathrm{keV}}\approx 10^{35}\, \rm erg\, \rm s^{-1}$. 
%
This represents the fraction of the stellar wind mechanical luminosity that is not lost by, e.g., forbidden-line emission of the hot diluted gas in the 
shock wind region and/or by the ISM gas heated by thermal transfer~\citep{mackey_sept_2014,2019A&A...625A...4G}. 
The lightcurves in the other energy bands exhibit similar behaviour, albeit offset by 1, 2 and 4 order of magnitude for the 
luminosities $L_{X}^{\mathrm{E}\ge0.5\mathrm{keV}}$, $L_{X}^{\mathrm{E}\ge1.0\mathrm{keV}}$ and $L_{X}^{\mathrm{E}\ge5.0\mathrm{keV}}$, respectively. The \mpo{low level} of $L_{X}^{\mathrm{E}\ge5.0\mathrm{keV}}$ illustrates 
that the hot gas ($T\ge 10^{7}\, \rm K$) emitting by free-free emission \mpo{negligibly} contributes to cooling in the nebula. 
The soft X-ray luminosity that we obtain from our model is consistent with predictions for nebulae around Wolf-Rayet stars, i.e. of 
the order of $10^{33}\, \rm erg\, \rm s^{-1}$~\citep{Dwarkadas_HEDP_2013,toala_aa_559_2013,toala_aj_147_2014}, and 
the hard component from hot gas is dimmer~\citep{chu_apj_599_2003}. 
This trend is not greatly affected by the stellar motion of the high-mass star, see Fig.~\ref{fig:luminosities}b-d, except during 
the post-main-sequence evolutionary phases during which the highest-energy emission, $L_{X}^{\mathrm{E}\ge5.0\mathrm{keV}}$ ,
exhibits stronger variations as $v_{\star}$ increases, reflecting the fact that the wind material released by the star interacts differently with the the main-sequence bubble an/or the ISM material, see our model Run-40-CSM at times $3.2$$-$$3.7\, \rm Myr$ (Fig.~\ref{fig:luminosities}d).

The post-supernova evolution of the thermal X-ray luminosities \mpo{begins with} $L_{X}^{\mathrm{E}\ge0.1\mathrm{keV}}\approx 10^{34}\, \rm erg\, \rm s^{-1}$ in the static star case (Fig.~\ref{fig:luminosities}e) and slightly more for a runaway progenitor (Fig.~\ref{fig:luminosities}h). 
\dmam{
A sudden rise of the lightcurves happens when the forward shock collides with the wind termination shock. In the static case this happens around $25\, \rm kyr$ after the explosion (see Fig.~\ref{fig:luminosities}e and Fig.~\ref{fig:snr}a,d) 
and sooner if the bulk motion of the progenitor is larger (Fig.~\ref{fig:luminosities}f-h) since the stand-off distance 
of the corresponding bow shock is smaller~\citep{wilkin_459_apj_1996}. The luminosities then decrease, nonetheless, they are interspersed with additional peaks and/or variations provoked by the reflections of the reverberated shock wave onto 
the cavity's wall. 
} 
As before the time of the explosion, the luminosities in the more energetic bands are about $1$, $2$ and $3$ order of magnitude 
lower for $L_{X}^{\mathrm{E}\, \ge0.5\, \mathrm{keV}}$, $L_{X}^{\mathrm{E}\, \ge\, 1.0\mathrm{keV}}$ and $L_{X}^{\mathrm{E}\ge5.0\mathrm{keV}}$, respectively, i.e. the remnants of runaway progenitors are easier to observe in the soft X-ray band than in the hard X-ray band. 
This order relation persists throughout the expansion of the supernova remnant in the ISM (Fig.~\ref{fig:luminosities}f-h). 
However, the hard X-ray emission decreases faster as function of time because the supernova remnant expands into the ISM and 
cools so that the hot gas contribution to the emission is less and less important~\citep{meyer_mnras_450_2015}.
%

Fig.~\ref{fig:luminosities_tracers} plots the evolution of the stellar wind  contribution to the thermal X-ray luminosity, 
calculated for each energy band as, 
\begin{equation}
        L_{X}^{\textcolor{black}{\mathrm{E}\ge\alpha}} =  \iint_{\mathrm{SNR}} j_{\mathrm{X}}^{\textcolor{black}{\mathrm{E}\ge\alpha}}(T) n_{\mathrm{H}}^{2} Q_{\mathrm{1}}  \mathrm{dV},
        \label{eq:lum_x_sw}  
\end{equation}
where $Q_{\mathrm{1}}$ is the passive scalar discriminating stellar wind from other kinds of material. 
The $\mathrm{E}\ge5.0\mathrm{keV}$ luminosity mostly originates from the stellar wind while the emission at $\mathrm{E}\ge1.0\mathrm{keV}$ 
\mpo{largely} comes from the shocked ISM gas. 
\mpo{The stellar-wind contribution to the latter} is about a few per cent when the bubble grows at times $\le 1.5\, \rm Myr$ 
and increases up to $\ge 80\%$ when the star has evolved to the post-main-sequence phases, see our models Run-0-CSM and Run-10-CSM (Fig.~\ref{fig:luminosities_tracers}a,b). 
In the runaway progenitor cases this contribution reaches only about $ 60\%$ (Fig.~\ref{fig:luminosities_tracers}d) which means that the expelled luminous-blue-variable and Wolf-Rayet shells emit more by thermal Bremsstrahlung than the ISM material. 
\mpo{In the other energy bands the stellar wind contribution of their emission is negligible during 
the main-sequence phase, as it represents $\le 10$$-$$15\%$ of the overall emission, but it progressively increases}
when the bubble is formed at time $1.5\, \rm Myr$. 
This changes during the Luminous-Blue-Variable and the Wolf-Rayet phases when the wind contribution to $L_{X}^{\mathrm{E}\ge0.5\mathrm{keV}}$  
rises up to $90\%$. Note that for a static star $L_{X}^{\mathrm{E}\ge0.1\mathrm{keV}}$ exhibits variations 
reflecting the successive mass-loss rate variations happening during the luminous blue variable eruptions and Wolf-Rayet events  
(Fig.~\ref{fig:luminosities_tracers}a). 
These variations of the stellar-wind contribution to $L_{X}^{\mathrm{E}\ge0.1\mathrm{keV}}$ \mpo{are 
independent of the speed} of the progenitor star (Fig.~\ref{fig:luminosities_tracers}b-d). 
%

%
\textcolor{black}{
The proportion of thermal-X-ray-emitting wind material in the supernova remnants is a function of the 
bulk motion of the progenitor star, and, consequently, relates on the distribution of circumstellar material 
at the supernova time. 
Its largest contribution, about $10\%$, arises in the $E\ge5.0\mathrm{keV}$ energy band 
(Fig.~\ref{fig:luminosities_tracers}e-h).   
This further illustrates how the contribution of stellar wind to the soft X-ray emission in the remnant 
decreases with gas temperature. 
Once the expansion of the supernova shock wave in the unperturbed stellar wind ceases ($<5$-$10\, \rm kyr$), 
the wind contribution to the hard X-ray emission decreases from about $80\%$ in the static case to \mpon{below $10\%$ for} our fastest-moving progenitor. 
Our results highlight that thermal X-ray emission of remnants from fast-moving progenitors originates from ISM 
material, as most of the wind gas has been advected far from the location of the explosion and had time to cool 
before the moment of the blastwave-circumstellar medium interaction. 
\mpon{Note} that the figure represent the proportion of wind material exclusively, the rest originating from the sum of the shocked ISM 
gas plus the ejecta melting into the supernova remnant (see Section~\ref{sect:mixing}). 
The ejecta contribution to the thermal X-ray emission is not important compared to that of the wind and ISM because its mass is much smaller than the overall remnant mass. 
}

\subsection{Mixing of material}
\label{sect:mixing}

\begin{figure*}
        \centering
                \includegraphics[width=0.9\textwidth]{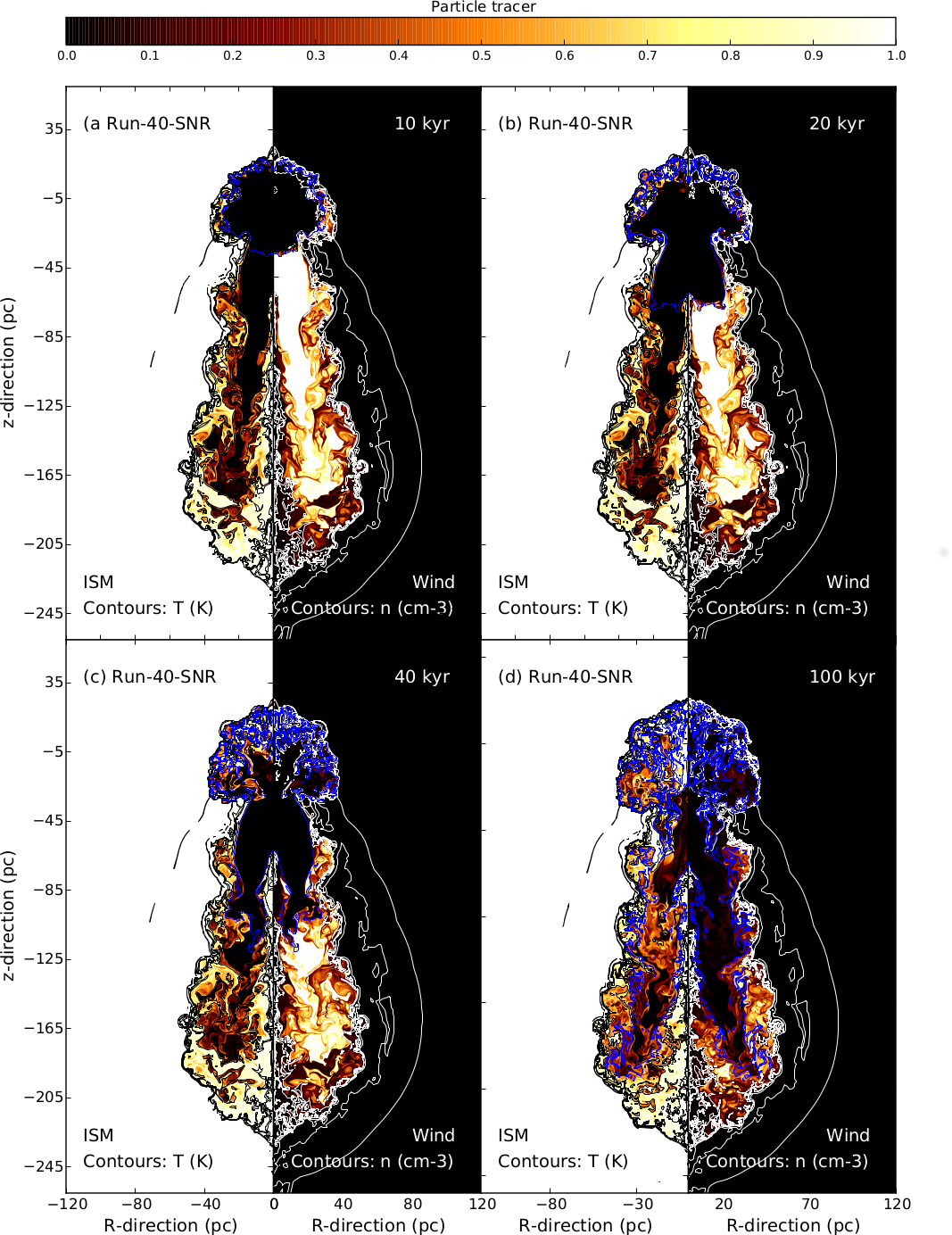}
        \caption{Same as Fig.~\ref{fig:snr_20kms_mixing} but here for a progenitor moving with velocity $40\, \rm km\, \rm s^{-1}$. 
                 \textcolor{black}{
                 The figures plot the value of the quantity \textcolor{black}{1-$Q_{1}$ representing the proportion of 
                 ISM material (left part of the panels) and the value of the passive scalar tracer $Q_{1}$ standing for the proportion of stellar wind gas 
                 (right part of the panels)}, respectively. The blue contours of the passive scalar field $Q_{2}$ indicate a 
                 $10\%$ contribution of ejecta in number density. 
                 }
        }      
        \label{fig:snr_40kms_mixing}  
\end{figure*}

Fig.~\ref{fig:snr_20kms_mixing} illustrates the temporal evolution of the mixing of ejecta/wind/ISM material in our supernova remnant model Run-20-SNR for 
a progenitor star moving with speed $v_{\star}=20\, \rm km\, \rm s^{-1}$. The panels correspond to the remnant at times 
$10$ (a), $20$ (b) $40$ (c) and $100\, \rm kyr$ (d) after the explosion, respectively. On each plot, the left-hand side shows the value of $1-Q_{1}$, 
i.e. the proportion of ISM gas in the computational domain and the right-hand side shows the value of the tracer $Q_{1}$ for the 
stellar wind material, respectively, both spanning from 0 to 1, with 0 standing for the complete absence of a given species while 1 stands for 
a medium exclusively made of this kind of species, respectively. The black contours indicate the locations where the remnant 
temperature is $T=10^{5}$, $10^{6}$, and $10^{7}\, \rm K$, while the white lines are iso-contours for the gas density with levels
$n=1.0$, $10^{1}$, $10^{2}$, and $10^{3}\, \rm cm^{-3}$. Additionally, the overplotted blue contours trace the regions at which ejecta contributes $10\%$ of the gas.

\textcolor{black}{
After $10\, \rm kyr$ the supernova blastwave bounces against the ring opened by the final Wolf-Rayet wind of 
the defunct star, where it is partially reverberated towards the center of the explosion and partially transmitted 
to the ISM through the bow shock according to the mechanism described in~\citet{borkowski_apj_400_1992,meyer_mnras_450_2015}. 
The supernova remnant is mostly filled with hot stellar-wind gas surrounded by a main-sequence shell of dense, 
cold ISM material. The interface between wind and ISM is the former unstable contact discontinuity of the 
pre-supernova bubble with which the \mpon{various} post-main-sequence shells collided (Fig.~\ref{fig:snr_20kms_mixing}a). 
}
After $20$ to $40\, \rm kyr$ the supernova shock wave that passed through the Wolf-Rayet wind keeps on 
expanding into the unperturbed ISM, and the reflection of its forward shock to the center of the explosion 
begins, enhancing the mixing of stellar wind and ISM material. This is the process which is responsible for 
the formation of Cygnus-loop supernova remnant, although that particular objects concerns a lower-mass, red supergiant progenitor~\citep{meyer_mnras_450_2015,fang_mnras_464_2017}.

\textcolor{black}{
Simultaneously, the reverberated shock wave develops instabilities at the interface between ejecta and mixed 
wind/ISM material (Fig.~\ref{fig:snr_20kms_mixing}b,c) while the region occupied by the ejecta further expands 
into the interior of the wind blown cavity (Fig.~\ref{fig:snr_20kms_mixing}c).
}
\textcolor{black}{
After $40\, \rm kyr$ the supernova remnant assumes an asymmetric morphology made of the large, spherical, 
former main-sequence bubble (lower part) and the Wolf-Rayet-produced cocoon (upper part). The bubble concentrates 
the stellar wind material surrounded by a layer of shocked ISM mixed with ejecta material while the almost 
ejecta-free upper mushroom is filled with a mixture of wind and ISM gas. 
The ejecta material lies in a tubular zone located at the inside of the bottom region of the remnant, directed 
towards to the direction of motion of the supernova progenitor, as well as in the region to the former center of 
the explosion (Fig.~\ref{fig:snr_20kms_mixing}d). 
}

\textcolor{black}{
Fig.~\ref{fig:snr_40kms_mixing} is similar to Fig.~\ref{fig:snr_20kms_mixing} for our model Run-40-SNR with a 
runaway progenitor star moving at $v_{\star}=40\, \rm km\, \rm s^{-1}$. 
As the progenitor rapidly moves, the main-sequence bubble is elongated. The blastwave bounces 
against the stellar wind bow shock of the progenitor which is separated from the nebula that was 
produced during its main-sequence phase of evolution. This advects the ejecta all across the Wolf-Rayet 
shell-like region generated during the progenitor's ultimate evolutionary stage, which is centered 
onto the location of the explosion (Fig.~\ref{fig:snr_40kms_mixing}a,b). 
}
\textcolor{black}{
Between \textcolor{black}{$20\, \rm kyr$} and $40\, \rm kyr$ the shock-wave/Wolf-Rayet-shell interaction continues, and \mpon{so does the mixing of 
material, as} part of the supernova shock wave expands into the low-density cavity of 
unperturbed wind material, along the direction opposite of the motion of the progenitor. 
Pushed downward by the ejecta, the former unperturbed shocked stellar wind develops instabilities 
at its interface with the shocked ISM (Fig.~\ref{fig:snr_40kms_mixing}b,c). 
}

\textcolor{black}{
At time $40\, \rm kyr$ the supernova remnant has an internal structure different from that of Run-20-SNR 
(with $v_{\star}=20\, \rm km\, \rm s^{-1}$). It is made of an upper quasi-circular region and a longer, 
tail-like structure. Both regions have efficiently mixed stellar wind and ISM gas. The ejecta are both 
located in the upper cocoon and in the main-sequence wind bubble (Fig.~\ref{fig:snr_40kms_mixing}d). 
Models Run-20-SNR and Run-40-SNR reveal the importance of progenitor motion in the mixing of different 
materials in remnants of massive runaway progenitors. The degree of wind/ISM mixing and the distances from the 
center of the explosion at which it happens increases with $v_{\star}$ (Fig.~\ref{fig:snr_20kms_mixing}d, Fig.~\ref{fig:snr_40kms_mixing}d) as a consequence of the elongation of the stellar wind bubble. 
It channels the supernova shock wave when the Mach number of the progenitor moving through the ambient 
medium is $v_{\star}/c_{\rm s}\sim2$, with $c_{\rm s}$ the unperturbed ISM sound speed, while the 
ejecta distribution remains is spherical only if the progenitor is either at rest or slowly-moving and explode 
inside and/or off-centered in its main-sequence circumstellar bubble, at least in the ambient medium that 
we consider. 
}


\section{Discussion}
\label{sect:discussion}

This section presents the limitations of our method, discusses our results in the context of previous studies, 
compares our results with observations of asymmetric supernova remnants, and further discusses the potential role 
of stellar wind nebula as cosmic-ray accelerators. 

\subsection{Caveats of the models}
\label{sect:caveats}

The first limitation of our method is the two-dimensional nature of the hydrodynamical simulations, which 
restricts the realism of the modelled wind bubbles and supernova remnants. Axisymmetry in \mpo{particular} prevents us from 
investigating the effects of, e.g., the angles between the direction of motion of the progenitor star, the local interstellar
magnetic field, and the axis of stellar rotation. 
Three-dimensional simulations such as in~\citet{katushkina_MNRAS_465_2017,katushkina_MNRAS_473_2018} would be highly \mpo{desirable for} a more realistic depiction of the surroundings of massive stars, however, their huge 
computational costs would \mpo{inhibit} scanning the parameter space. 
%
%
The treatment of the intrinsic nature of the ISM is simplified by assuming a uniform medium of number density 
$1\, \rm cm^{-3}$ and temperature $8000\, \rm K$~\citep{meyer_2014bb}. Any intrinsic native turbulence 
in the velocity field of the medium, clumps, or filamentary structures that could potentially influence 
the development of shocks wave are also neglected, see~\citet{gvaramadze_mnras_474_2018}. 
The stellar wind history used as boundary conditions is also not unique, and other evolutionary tracks \mpo{and alternative 
stellar rotation rates can change the luminous blue variable or 
the Wolf-Rayet phase of massive stars \citep{ekstroem_aa_537_2012}. They would produce a different wind bubble and consequently result in 
dissimilar} supernova remnants. 
%
%

\subsection{Comparison with previous studies}
\label{sect:other}

Our method uses a self-consistently calculated stellar evolution history as boundary for the stellar wind, and the 
hydrodynamical simulations are performed for a different corner of the parameter space and on a higher spatial grid 
resolution than in other studies. 
Previous works tackled the problem of the circumstellar medium of Wolf-Rayet-evolving 
stars either considering static stars~\citep{garciasegura_aa_455_1995a,garciasegura_1996_aa_316} 
or \mpo{using lower-resolution calculations of} moving stars in the context of the formation of dense circumstellar regions eventually leading to gamma-ray 
bursts~\citep{eldridge_mnras_367_2006,vanmarle_aa_460_2006,vanmarle_aa_469_2007,eldridge_mnras_414_2011}.   
The series of work of~\citet{brighenti_mnras_270_1994,brighenti_mnras_273_1995,brighenti_mnras_277_1995} is 
the closest to our approach in the sense that it explores the morphology of wind bubbles blown by 
massive stars by first calculating the evolution of their surroundings before launching a blastwave in it.

Only one study of this series investigates the shape of \textcolor{black}{supernova remnants from} moving Wolf-Rayet progenitors, altough not priorily undergoing 
luminous-blue-variable eruptions~\citep{brighenti_mnras_270_1994}. 
%
%
Their model~1 has ISM properties and a stellar motion similar to our run 
Run-20-SNR, however, assuming a different stellar wind history as it includes a cool red supergiant phase while we here simulate 
luminous blue variable and Wolf-Rayet stars. A slowly-moving progenitor (our calculations Run-10-CSM and Run-10-SNR) is also 
not considered in that work. 
Last, a couple of works have been tailored to the study of specific young supernova remnants, \mpo{whose shape 
might} be explained by the presence of a bow shock around the progenitor prior to its explosion. These studies 
mainly concern the Crab nebula~\citep{cox_mnras_250_1991} and the Kepler supernova 
remnant~\citep{borkowski_apj_400_1992,velazquez_apj_649_2006,chiotellis_aa_537_2012,toledo_mnras_442_2014}. 
These models \mpo{apply to much younger ($\sim 1000\, \rm yr$) supernova remnants than ours and cover a} different corner of the problem's parameter space.  

%
%
%

\subsection{Comparison with observations}
\label{sect:obs}

\subsubsection{\textcolor{black}{Comparison with existing wind nebulae: the case of BD+43$\degree$3654}}
\label{sect:sub_csm}

\textcolor{black}{
The most obvious object permitting a comparison with our models for the pre-supernova circumstellar medium of 
the $60\, \rm M_{\odot}$ of~\citet{groh_aa564_2014} is the very massive $\simeq 55$-$85\, \rm M_{\odot}$  
runaway star BD+43$\degree$3654, which probably escaped from the Cygnus OB2 region~\citep{comeron_aa_467_2007}. 
BD+43$\degree$3654 is an $\approx 1.6\, \rm Myr$-old, O4If-typed main-sequence stellar object stellar which 
properties have been constrained to be $\dot{M}\approx 10^{-5}\, \rm M_{\odot}\, \rm yr^{-1}$ and 
$v_{\rm w}\approx 2300\, \rm km\, \rm s^{-1}$, respectively. Its surrounding is shaped as a well-defined, 
axisymmetric bow-shock nebula visible in the infrared. 
Using the observed geometry of the bow shock, its proper motion has been estimated to be 
$\approx 40\, \rm km\, \rm s^{-1}$ and its ambient medium density to be $n_{\rm ISM}=6\, \rm cm^{-3}$~\citep{comeron_aa_467_2007}. 
}

\textcolor{black}{
Nevertheless, the formula for the stand-off distance of a bow shock measured along the direction of motion of its driving star,
\begin{equation}
	R_{\rm SO} = \sqrt{\frac{\dot{M}v_{\mathrm{w}}}{4 n_{\mathrm{ISM}}v_{\star}^{2} } },
    \label{eq:wilkin}
\end{equation}
informs that $n_{\rm ISM}$ is a function of $\dot{M}$~\citep{baranov_sphd_15_1971,wilkin_459_apj_1996} 
and the discoverers of BD+43$\degree$3654's bow shock discussed in ~\citet{comeron_aa_467_2007} that 
any overestimate of the wind properties $\dot{M}$ or $v_{\rm w}$ will translate into a overestimate 
of its background ISM density, which may therefore be smaller than $n_{\rm ISM}=6\, \rm cm^{-3}$ by a factor 
of $\approx 3$. This is comparable with the value of $\simeq 0.79\, \rm cm^{-3}$ used in our study. 
By interpolating our stellar evolutionary track, we find $\dot{M}\approx 10^{-5}\, \rm M_{\odot}\, \rm yr^{-1}$ and $v_{\rm w}\approx 2865\, \rm km\, \rm s^{-1}$ for the stellar age of BD+43$\degree$3654 $\approx 1.6\, \rm Myr$, that 
is rather consistent with the derivations of~\citet{comeron_aa_467_2007}. Hence, our model Run-40-CSM fits well  
the case of BD+43$\degree$3654 and our simulation can be considered as a prediction of the future evolution of its pre- and eventual post-supernova circumstellar medium. 
}

\subsubsection{\textcolor{black}{Comparison with existing supernova remnants}}
\label{sect:sub_snr}

We note that~\citet{brighenti_mnras_270_1994,brighenti_mnras_277_1995} proposed that barrel-shaped supernova remnants 
such as G296.5+10.0~\citep{storey_aa_265_1992,harvey_apj_712_2010} and other bipolar remnants of similar 
morphology~\citep{kogan_apss_166_1990} have been generated by the interaction between a supernova shock wave with 
the walls of a low-density wind cavity produced by the stellar motion of its own progenitor, into which the forward shock 
has been channeled~\citep{cox_mnras_250_1991}. 
This idea has been supported by~\citet{meyer_mnras_450_2015} with optical H$\alpha$ and [O{\sc iii}] synthetic emission maps derived from 
hydrodynamical simulations of red-supergiant-evolving runaway progenitor stars. However, most such supernova remnants have been shown to be aligned with the galactic magnetic field and directed towards the Galactic center. This observational 
evidence \mpo{suggests that their formation is better explained as effect of the interstellar magnetic field in} 
Galactic spiral arms \citep{gaensler_apj_493_1998}. 
This hypothesis has been \mpo{supported} with both magneto-hydrodynamical simulations and non-thermal synchrotron and radio emission maps~\citep{orlando_aa_470_2007}. 

We can also see noticeable differences between our supernova remnants models generated by runaway luminous blue variable/Wolf-Rayet 
progenitors and those for runaway red supergiant progenitors \citep[see][]{meyer_mnras_450_2015}. 
In the case of a red supergiant progenitor, the ejecta blows out beyond the last wind bow shock and freely expands into the ISM, if the 
progenitor moves sufficiently fast and/or if the stellar wind is not too dense. A heavier ZAMS stars and/or a slower progenitor 
generate a denser bow shock of swept-up ISM which will trap the new-born cold red supergiant shell and the subsequent shock wave. \mpo{This feature is present in all of our runaway ($v_{\star} \ge 20\, \rm km\, \rm s^{-1}$) Wolf-Rayet progenitor runs}. 
One should therefore expect to observe remnant from runaway Wolf-Rayet progenitor with Cygnus-Loop-nebula-like morphology, 
which \mpo{would resemble a} remnant of a runaway red supergiant-evolving 
star~\citep{tenoriotagle_aa_148_1985,aschenbach_aa_341_1999,meyer_mnras_450_2015,fang_mnras_464_2017}, although of different 
size and chemical composition. 
Inversely, our results \mpo{argue against a $ 60\, \rm M_{\odot}$ 
ZAMS star as progenitor of} historical supernova remnants such as Kepler's or Tycho and support solutions involving symbiotic binary systems of a runaway AGB star together with an type Ia explosive 
companion~\citep{borkowski_apj_400_1992,velazquez_apj_649_2006,vigh_apj_727_2011,chiotellis_aa_537_2012,williams_apj_770_2013,toledo_mnras_442_2014}.


%
The supernova remnant G109.1-1.0 (CTB 109) has for long been suspected to be an isolated supernova remnant whose progenitor 
may have moved quickly \citep{brighenti_mnras_270_1994,brighenti_mnras_277_1995}. However, evidence of dense 
molecular material aside of the remnant~\citep{cruces_mnras_473_2018} suggests it being produced by a shock wave evolving 
into a rather low-density medium that is sparsed with higher-density cloudlets~\citep{coe_mnras_238_1989,sasaki_apj_642_2006}, 
see also the numerical simulations of~\citet{bolte_aa_582_2015}.  
The remnant VRO 42.05.01~\citep{arias_aa_622_2019} may be the archetypal example of a supernova remnant interacting with an ISM cavity. 
The low-density region inducing the asymmetries may have been formed by a previous supernova~\citep{rho_apj_484_1997} or by the motion of the 
progenitor star itself~\citep{2019arXiv190906131D}. Recent observations of molecular emission and numerical studies tailored to VRO 42.05.01 
suggest that the progenitor could have been a massive runaway star interacting with a hot diluted ISM tunnel~\citep{2019arXiv190906131D}, 
but \mpo{indicated also} that an asymmetric Wolf-Rayet wind is sufficient to produce its overall morphology~\citep{2019arXiv190908947C}. 
The influence of the anisotropy of post-main-sequence stellar winds should be investigated in future studies. 
\mpo{The supernova remnant S147 has been suggested to result from a Wolf-Rayet 
progenitor~\citep{gvaramadze_aa_454_2006} in a} multiple system~\citep{dincel_mnras_448_2015}. 

\subsection{ Are Wolf-Rayet nebulae efficient cosmic rays accelerators ? }
\label{sect:crs}

There is a growing interest in the non-thermal surroundings of massive stars. 
The strong winds of OB and Wolf-Rayet stars generate magnetised termination shocks such that a proportion of the stellar mechanical luminosity 
\mpo{could power high-energy} cosmic rays~\citep{webb_apj_298_1985,seo_jkas_51_2018}. 
A number of numerical simulations calculating the acceleration and diffusion of cosmic rays in \mpo{static stellar-wind/bow-shock environments} of massive stars have been 
performed~\citep{delvalle_aa_550_2013,delvalle_mnras_448_2015,delvalle_apj_864_2018}, and speculative predictions were announced for the Bubble 
nebula, a runaway OB star moving through dense molecular gas \citep{2019A&A...625A...4G}. 
Particle acceleration \mpo{and the production of
non-thermal emission in the surroundings of pre-supernova (runaway) massive stars are expected to arise from processes similar to those operating} in young supernova 
remnants~\citep{reynolds_apss_336_2011}, albeit with smaller efficiency~\citep{voelk_apj_253_1982,zirakashvili_aph_98_2018}. 

\mpo{Per object,} circumstellar cosmic-ray feedback of high-mass stars is of lower importance in the Galactic 
energy budget compare to other non-thermal-emitting objects such as supernova remnants~\citep{seo_jkas_51_2018,2019arXiv190901424R} or \textcolor{black}{colliding winds} in 
binaries~\citep{benaglia_aa_399_2003,debecker_mnras_371_2006,2006ApJ...644.1118R,debecker_aa_558_2013}. 
\mpo{Accordingly non-thermal X-rays from stellar-wind bow shocks surrounding OB 
stars have not been detected to date \citep{toala_apj_821_2016,DeBecker_mnras_471_2017,tola_apj_838_2017,binder_aj_157_2019}. 
\citet{2019arXiv190912332P} recently claimed the detection of} synchrotron emission from a series of shells and arc around the 
Wolf-Rayet nebula G2.4+1.4. \mpo{Our results further explore the circumstellar medium of evolved massive stars and provide the background on which we shall investigate} the synchrotron, inverse Compton and Gamma-ray emission, 
e.g. using the {\sc ratpac} code~\citep{telezhinsky_aa_541_2012,telezhinsky_aph_35_2012,telezhinsky_aa_552_2013}. 


\section{Conclusion}
\label{section:cc}

\textcolor{black}{
Motivated by the existence of stars, whose mass exceeds $\sim60\, \rm M_{\odot}$ and which can 
either be affected by high proper motion and/or generate circumstellar nebulae by stellar 
wind-ISM interaction~\citep{2010ApJ...714L..26T,2011MNRAS.410..304G,2013MNRAS.430L..20G}, 
we investigate the shaping of circumstellar nebulae around a non-rotating $60\, \rm M_{\odot}$ 
stars \citep{groh_aa564_2014} and we explore how they can influence the development of 
asymmetries in their supernova remnants. 
}
These stars are characterised by a violent evolution history including several successive 
luminous-blue-variable and Wolf-Rayet phases. 
Using high-resolution 2D hydrodynamical simulations performed with the {\sc pluto} 
code~\citep{mignone_apj_170_2007, migmone_apjs_198_2012}, we investigate how the ejection of 
the dense shells associated to these evolutionary phases can couple to the stellar 
proper motion and carve aspherical low-density cavities of wind material, inside of which 
the star dies as a core-collapse supernova. The resulting remnant therefore adopts anisotropies 
reflecting the imprint of star's past evolution history onto its close 
surroundings~\citep{meyer_mnras_450_2015}. 
We consider several static, slowly-moving, and fast stars evolving in the warm phase of the ISM, 
and we predict the thermal X-ray signature of both their stellar wind nebulae and supernova remnants 
up to $100\, \rm kyr$ after the explosion.

\textcolor{black}{
We show that, as long as such a massive ($\sim 60\, \rm M_{\odot}$)  progenitor star moves 
supersonically through the ISM, its subsequent supernova remnant will \mpon{display similarity} to that of the Cygnus-loop nebulae.  
The role of the strong luminous-blue-variable and Wolf-Rayet shells winds predicted by the Geneva 
stellar evolutionary tracks~\citep{groh_aa564_2014} add to the asymmetries produced by the stellar 
motion as they induce outflows that can pierce the main-sequence wind bubble of the progenitor star, 
opening a route for the later supernova shock wave. 
A rather modest bulk motion ($10\, \rm km\, \rm s^{-1}$) of our $60\, \rm M_{\odot}$ star 
is not sufficient to break the sphericity of the overall remnant as the expanding supernova shock wave 
is trapped into the defunct, quasi-spherical stellar wind bubble~\citep{weaver_apj_218_1977}. 
}

\dmam{
However, as the progenitor star moves faster, the explosion is off-set with respect to the geometrical center 
of the wind nebula and, in the runaway limit of the stellar motion, happens outside of the bubble, 
in a ring of Wolf-Rayet and ISM material. 
The shock wave blows out of the wind nebula, resulting in an homogeneous mixing of material throughout 
the elongated remnant, while ejecta is reverberated around the center of the explosion.
This leads to the formation of two-lobes structures ahead and behind the center of the explosion, 
inside of which the degree of mixing of ejecta with wind and ISM materials is correlated to the 
progenitor's motion.
Our pre-supernova nebulae and supernova remnants are bright in soft thermal X-rays from shocked ISM 
gas, except during the luminous-blue-variable eruptions of the progenitor when it comes from the stellar 
wind. Additionally, such remnants of low-velocity progenitors exhibit a hard X-rays component of shocked stellar wind.  
}

\textcolor{black}{
Our results stress that a single runaway high-mass star can affect the chemical composition of 
an entire local region of the ISM, see e.g.~\citet{2010ApJ...714L..26T}. 
Consequently, we highlight the importance of chemically investigating the large-scale surroundings 
($\sim 100\, \rm pc$) of supernova remnants, as it is a trace of the possible runaway nature of the 
defunct star(s), but also since the distribution of enriched material that has mixed in it with the 
ISM informs on the past stellar evolution of supernova progenitor(s).
}
Furthermore, our simulated remnants indicate that luminous-blue-variable/Wolf-Rayet-type progenitors 
can not be the origin of \textcolor{black}{supernova remnants with morphologies such as those of} Cas~A, RCW86, Kepler or Tycho. Their young age ($\le 2500\, \rm yr$) and compact size are not in accordance with our results, which either support scenarios 
involving type Ia supernova and/or suggest a non-Wolf-Rayet-evolving core-collapse progenitor. 
Stellar wind-ISM interaction should therefore be further investigated by scanning the parameter study of the problem 
and by modelling such remnants with, e.g. global 3D magneto-hydrodynamical simulations. 
Particularly, the large termination shocks which form around luminous-blue-variable and Wolf-Rayet stars suggest that 
these wind nebulae might accelerate particles as high-energy cosmic rays. This is consistent with the recent monitoring 
of synchrotron emission from the Wolf-Rayet shell nebula G2.4+1.4~\citep{2019arXiv190912332P} and motivates future 
observational and theoretical studies on the non-thermal emission of circumstellar structures such as synchrotron and 
inverse Compton radiation.


\section*{Acknowledgements}

\textcolor{black}{The authors thank the anonymous referee for comments which improved the quality of the paper.}
DMAM and M.~Pohl acknowledge B.~Hnatyk for constructive discussion.
The authors acknowledge the North-German Supercomputing Alliance (HLRN) for providing HPC 
resources that have contributed to the research results reported in this paper. 
M.~Petrov acknowledges the Max Planck Computing and Data Facility (MPCDF) for providing 
data storage resources and HPC resources which contributed to test and optimise the {\sc pluto} code. 
This research made use of the {\sc pluto} code developed at the University of Torino  
by A.~Mignone and collaborators, the Matplotlib plotting library for the Python programming language 
and the {\sc xspec} X-ray spectral fitting package developed by K.~Arnaud and collaborators at the University of Maryland.  
DMAM thanks the Lorentz Center at the University of Leiden for organising and hosting the workshop 
"Historical Supernovae, Novae and Other Transient Events" in October 2019, where many insightful discussions 
which participated to the completion of this paper took place. 
%


\bibliographystyle{mn2e}

\footnotesize{
\bibliography{grid}
}


\end{document}